\documentclass[aps,prx,twocolumn,superscriptaddress,longbibliography,floatfix]{revtex4-2}

\usepackage{amsmath}
\usepackage{amssymb}
\usepackage{bm}
\usepackage{dcolumn}
\usepackage{graphicx}
\usepackage{microtype}
\usepackage{float}
\usepackage[colorlinks=true,citecolor=blue,urlcolor=blue,linkcolor=blue]{hyperref}

\newcommand{\pmax}{p_{\mathrm{max}}}
\newcommand{\GammaPeak}{\Gamma_{\mathrm{peak}}}
\newcommand{\Iabs}{I_{\mathrm{abs}}}
\newcommand{\zpeak}{z_{\mathrm{peak}}}
\newcommand{\E}{\mathbb{E}}
\newcommand{\Prob}{\mathbb{P}}
\newcommand{\dd}{\mathrm{d}}
\newcommand{\summarycell}[2]{\parbox[t]{#1}{\raggedright #2\strut}}

\begin{document}



\title{Predictive Structure Behind Rare Outcomes in Random Quantum Circuits}

\author{Myeongsu Kim}
\affiliation{Department of Computer Science, Purdue University, West Lafayette, Indiana, USA}

\author{Travis Humble}
\affiliation{Quantum Science Center, Oak Ridge National Laboratory, Oak Ridge, Tennessee 37831, USA}

\author{Sabre Kais}
\email{skais@ncsu.edu}
\affiliation{Department of Electrical and Computer Engineering, North Carolina State University, Raleigh, North Carolina 27695, USA}
\affiliation{Department of Chemistry, North Carolina State University, Raleigh, North Carolina 27695, USA}

\date{\today}

\begin{abstract}
Local random quantum circuits can occasionally produce unusually large output probabilities. However, the final output distribution containing such a peak does not by itself reveal how the peak formed during circuit evolution. We investigate whether states formed during circuit evolution carry structural signatures that predict subsequent peak formation. In numerical experiments, circuits selected for large terminal peaks exhibit stronger pairwise correlations and probability redistribution, together with lower bipartite entanglement and output entropy, than controls generated by the same random-circuit model. Their intermediate states retain higher fresh-peak event probability when the remaining gates are independently regenerated. Within an unconditioned ensemble, a fixed structural score computed from the initial circuit segment predicts new-peak probability. Averaged equally across $n=8$--$16$, a replay with equal numbers of remaining-circuit evaluations shows that score-based allocation yields $1.80$ times as many peaks as uniform allocation and about $18\%$ more than a fixed baseline based on probability concentration and redistribution. Randomizing relative phases while preserving computational-basis probabilities strongly attenuates the selected-state advantage, demonstrating a functional contribution from relative-phase organization. Controlled gate-selection tests further show that the identified structure can guide circuit construction to promote peak formation. Together, these results connect rare output events to intermediate-state structure with predictive and constructive value under finite-depth local dynamics.
\end{abstract}

\maketitle


\section{Introduction}
\label{sec:introduction}

Local random quantum circuits provide a controlled setting for scrambling, entanglement growth, approximate unitary designs, and computational-basis output statistics \cite{HarrowLow2009,BrandaoHarrowHorodecki2016,Nahum2017,Nahum2018}. Their output statistics underpin sampling-based demonstrations of quantum computational advantage \cite{Arute2019}. Typical outputs are described through Porter--Thomas-like statistics and anticoncentration \cite{Boixo2018,Dalzell2022}, whereas the largest output probability is an extreme-value observable \cite{Lakshminarayan2008}.

Recent peaked-circuit studies develop verification protocols and explicit constructions, error-correction-based proposals, heuristic hardware demonstrations, and analyses of complexity and classical simulability \cite{AaronsonZhang2024,Zhang2025,DeshpandeEtAl2025,GharibyanEtAl2025,BravyiGossetLiu2024}. Kremer and Dupuis subsequently reported efficient classical simulation of the heuristic circuits used in Gharibyan et al.'s demonstration \cite{KremerDupuis2026}.

Alongside the construction and verification of peaked circuits, we investigate which intermediate-state properties predict new peak formation. To identify these properties, we compare circuits selected for unusually large terminal peaks (random-peaked, RP) with ordinary random reference circuits (random-normal, RN) generated using the same local random-circuit model. Conditioning on a terminal event reweights an entire trajectory ensemble \cite{GarrahanLesanovsky2010,ChetriteTouchette2015,Carollo2018}. We test whether the selected structure predicts new peak formation when the original remaining gates are replaced by independently drawn local circuits. Within an unconditional ensemble, we further ask whether combining structural observables improves prefix selection beyond fixed rules based on probability concentration and redistribution.

\subsection{Main results and scope}

First, a frozen prefix score predicts new-peak probability within an unconditioned ensemble and improves peak yield over uniform allocation and fixed probability-based selection rules at equal continuation-evaluation counts. Second, population-preserving phase intervention demonstrates a functional contribution from relative-phase organization. Third, gate-selection comparisons show that structural guidance improves high-peak yield under common refinement and new-peak probability under independent continuations.

The study uses exact, noiseless simulations of open one-dimensional circuits.

\section{Circuit ensembles and test overview}
\label{sec:setup}

We draw independent Haar-$U(4)$ gates on nearest-neighbor matchings of an open chain. A depth-$n$ prefix is followed by the reverse dagger of an independently drawn forward suffix of depth $n/2-\delta$, giving total nominal depth $L_\delta=3n/2-\delta$, with $\delta\in\{0,1,2,3\}$. The suffix is independent of the prefix before selection; it is not the inverse of that prefix. The analysis cut $t_*=n$ is fixed without using terminal outcomes. Along the primary-depth family ($\delta=0$), increasing $n$ increases both the prefix depth $n$ and continuation depth $n/2$. Primary-depth structural, continuation, prefix-prediction, and phase-intervention tests cover all five sizes, $n=8,10,12,14,16$. To assess robustness to total depth, we additionally test three shorter nominal depths at $n=8,10,12,14$. The separate construction and feature-ablation experiments use all four nominal depths at these four sizes. Some nominal-depth settings share the same terminal-unitary distribution, while their state--continuation distributions at the fixed cut can differ (Appendix~\ref{subsec:random-generator}). Full generator and cohort definitions are in Appendix~\ref{app:analysis-methods}.

For $U|0^n\rangle$, write $p_U(z)=|\langle z|U|0^n\rangle|^2$, $\pmax(U)=\max_z p_U(z)$, and $\GammaPeak(U)=2^n\pmax(U)$. Here $\GammaPeak$ measures the maximum output probability relative to the uniform value $2^{-n}$. A threshold $\gamma$ therefore corresponds to $\pmax(U)\geq\gamma/2^n$, with $\gamma_n^{\mathrm{RP}}$ specifying the structural RP selection threshold.

Structural observables do not enter cohort selection. Primary-depth structural cohorts contain $1000$ RN and $1000$ RP circuits at each of the five sizes. The additional-depth cohorts at $n=8,10,12,14$ use the same sample counts.

For an intermediate state $|\Psi_t\rangle$ and an independently drawn local continuation $V$, define the fresh-peak susceptibility
\begin{equation}
q_{\gamma,t}(\Psi_t)
=\Prob_V\!\left[\GammaPeak(V|\Psi_t\rangle)\geq\gamma\mid\Psi_t\right],
\label{eq:intro-susceptibility}
\end{equation}
where $q_{\gamma,t}$ is the probability that regenerated remaining gates produce a peak at or above $\gamma$ for the fixed state. A global-Haar continuation makes it independent of the input state by unitary invariance; finite-depth local continuations can retain sensitivity to intermediate-state organization. Table~\ref{tab:threshold-roles} distinguishes selection, fresh evaluation, and refinement targets.

\begin{table}[htbp]
\caption{\label{tab:threshold-roles} Threshold roles. $\gamma_n^{\mathrm{RP}}$ defines structural-cohort RP eligibility and high-peak construction evaluation; $\gamma_n$ defines fresh events and feature-ablation outcomes; $\gamma_n^{\mathrm{search}}$ is the refinement target. Thresholds are rounded here; calculations use unrounded values. The construction experiments use $n=8,10,12,14$.}
\begin{ruledtabular}
\begin{tabular}{cccc}
$n$ & $\gamma_n^{\mathrm{RP}}$ & $\gamma_n$ & $\gamma_n^{\mathrm{search}}$\\
\hline
8 & 30.0 & 16.09 & 30.0\\
10 & 31.5 & 17.28 & 31.5\\
12 & 33.2 & 19.55 & 33.0\\
14 & 35.1 & 19.41 & 35.1\\
16 & 36.4 & 20.73 & ---\\
\end{tabular}
\end{ruledtabular}
\end{table}

The continuation test discards each retained state's original suffix and applies common, independently drawn fresh banks to RP states and random RN controls. Events are counted at the terminal output of each complete continuation. The thresholds $\gamma_n$ are independently calibrated at the primary-depth empirical $0.999$ quantile and held fixed across depths. This calibration targets a common baseline event probability of approximately $10^{-3}$ across sizes at the primary depth; holding each threshold fixed allows that probability to vary at other depths. Fresh evaluation therefore tests a less extreme tail event than the original RP retention criterion.

An unconditional ensemble tests prediction using a fixed score computed from seven observables available by the analysis cut: all-pair and adjacent-pair quantum mutual information (QMI), redistribution at the last prefix transition, normalized Shannon entropy, collision ratio, normalized mean cut entanglement, and mean leading Schmidt weight. Each component is standardized using an independent, size-specific reference ensemble and oriented so that larger values follow the RP-associated structural pattern. Their equal-weight mean is standardized once more against the reference ensemble to define the prefix score. The observable types, orientations, and equal-weight rule are the same across sizes; size-specific reference scales are fixed before fresh-continuation evaluation, with no weights fitted to the evaluated outcomes.

A separate trajectory classifier tests cross-size ranking; it can use pre-final checkpoints after the analysis cut and is not a prefix predictor. Observable definitions and score specifications appear in Appendices~\ref{app:analysis-methods} and \ref{app:score-statistics}.

The phase test randomizes relative phases in pure-state replicas while preserving every computational-basis population. The construction tests select newly drawn gate candidates using size- and depth-matched RP observable targets. Trajectory and Haar initializations receive the same refinement budget; a separate greedy-tracker ablation compares multivariate with peak-only intermediate guidance at fixed endpoint targets. Figures~\ref{fig:selected-law}--\ref{fig:construction} report these tests. Continuation intervals account for shared states and suffixes while preserving paired comparisons; intervals across the size--depth grid are pointwise (Appendices~\ref{app:score-statistics} and \ref{app:native-depthgrid}).

\section{Results}
\label{sec:results}

\subsection{Susceptibility under independent continuations}
\label{subsec:susceptibility-results}

\label{subsec:inverse-design-results}

\begin{figure*}[t]
\centering
\includegraphics[width=0.94\textwidth]{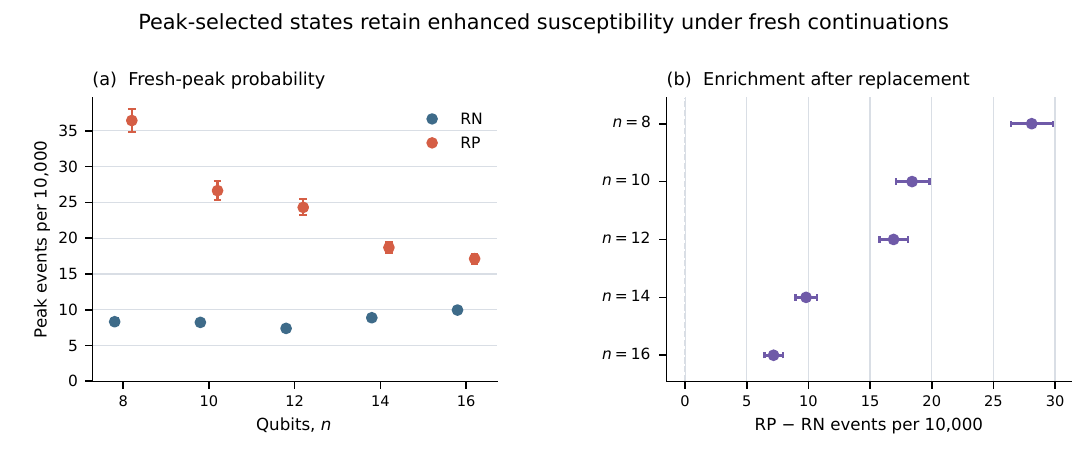}
\caption{\label{fig:selected-law} \textbf{Peak-selected states retain enhanced susceptibility after continuation replacement.} Original continuations are discarded before RN and RP receive common fresh banks, evaluated at the calibrated thresholds $\gamma_n$ in Table~\ref{tab:threshold-roles}. (a) Fresh-event probabilities and (b) their absolute RP-minus-RN differences at $n=8$--$16$. Rates count complete continuations exceeding the size-calibrated peak threshold, rather than the probability of an individual output bit string. Bars are $95\%$ intervals from resampling states and continuation columns while preserving shared-bank pairing.}
\end{figure*}

After discarding each original suffix, we evaluate RP and RN intermediate states using common banks of independently regenerated remaining circuits. At all five sizes, $n=8$--$16$, the fresh-peak event probability, averaged over states within each group, is higher for RP than for RN. Events use the independently calibrated thresholds $\gamma_n$, below the original high-peak selection criteria. Averaging the RP and RN event probabilities separately with equal weight per size and then taking their ratio gives $2.880$ [$95\%$ CI: $2.798$--$2.963$]. Figure~\ref{fig:selected-law} shows the size-specific event probabilities and risk differences; RN rates remain close to $10^{-3}$ across sizes under the calibrated event definition.

The RP advantage persists across all $12$ additional size--depth settings, with RP-to-RN risk ratios of $2.01$--$7.15$ and every $95\%$ interval above one (Appendix Fig.~\ref{fig:native-depthgrid}a). These are setting-specific effects at unchanged $\gamma_n$, rather than an across-size average. Together with the primary results, they establish enhanced peak probability under fresh continuations throughout the tested size--depth grid.

Terminal selection reweights the intermediate-state ensemble according to each state's probability of satisfying the original selection rule (Appendix~\ref{app:selection}). The tests below show that intermediate-state observables predict fresh-peak susceptibility within an unconditional ensemble.

A complementary repeatability test at $n=8,10,12,14$ finds positive covariance between state hit-rate estimates from independent row-specific banks within the unconditional parent ensemble, establishing repeatable between-state susceptibility differences without using the prefix score (Appendix~\ref{app:repeatability}).

\begin{figure*}[t]
\centering
\includegraphics[width=0.94\textwidth]{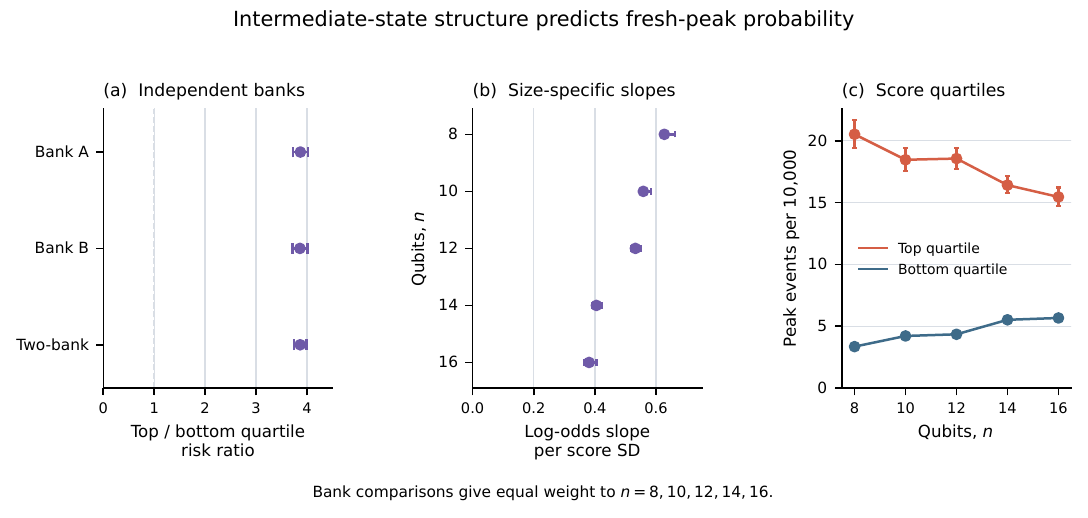}
\caption{\label{fig:susceptibility} \textbf{Intermediate-state structure predicts peak probability under regenerated continuations.} Unconditional cohorts at $n=8$--$16$ are evaluated at the calibrated thresholds in Table~\ref{tab:threshold-roles}. (a) Upper-to-lower score-quartile risk ratios for each independent bank and their combination; probabilities are averaged equally across all five sizes before forming each ratio. (b) Log-odds slopes per size-specific reference-score standard deviation. (c) Event probabilities in the reference-defined upper and lower score quartiles. Bars are $95\%$ state--continuation bootstrap intervals.}
\end{figure*}

Across $n=8$--$16$, the frozen seven-component prefix score predicts between-state differences in fresh-peak probability within unconditional ensembles. Averaging event probabilities equally across all five sizes gives an upper-to-lower score-quartile risk ratio of $3.868$ [$95\%$ CI: $3.744$--$3.993$]. The relationship is consistent across the two independent continuation banks, and size-specific log-odds slopes are positive throughout $n=8$--$16$ (Fig.~\ref{fig:susceptibility}). Because both the score and cohort are defined without the regenerated outcomes, this relation is prospective with respect to each fresh continuation.

In the primary frozen-bank replay across $n=8$--$16$, allocating the same number of continuation evaluations to the reference-defined upper score quartile instead of uniformly over all prefixes increases peak yield $1.80$-fold (Appendix Table~\ref{tab:budget-replay}). To assess the value of the combined structure beyond simpler selection rules, we compare it with high current prefix peak, low Shannon entropy, high collision ratio, and a fixed probability/history (PH) score combining redistribution, entropy, and collision. Each ranked baseline selects the same number of prefixes as the full score and uses the same continuation banks (Appendix~\ref{app:baseline-policy-comparison}). Relative to PH, the full score yields $2.761$ additional peaks per $10\,000$ evaluations [$95\%$ CI: $2.527$--$3.010$], corresponding to a yield ratio of $1.183$ [$95\%$ CI: $1.165$--$1.201$]. Table~\ref{tab:prefix-baselines} reports all fixed selection rules.

\begin{table}[t]
\caption{\label{tab:prefix-baselines} \textbf{Structural selection improves yield over probability-based baselines.} Primary-depth replay on unconditional prefixes at $n=8$--$16$. Event probabilities are averaged equally across banks and then sizes; peaks per $10\,000$ are expected counts at a common evaluation budget. All ranked rules use the full score's reference-quartile selection count within each size. The $95\%$ intervals are from the paired comparison in Appendix~\ref{app:baseline-policy-comparison}.}
\small
\setlength{\tabcolsep}{3pt}
\begin{ruledtabular}
\begin{tabular}{@{}lrr@{}}
Selection rule & Peaks per & Yield / uniform\\
 & $10\,000$ & [$95\%$ CI]\\
\hline
Uniform & 9.956 & 1.000\\
High prefix peak & 12.719 & 1.278 [1.247--1.308]\\
Low entropy & 14.982 & 1.505 [1.475--1.535]\\
High collision & 14.732 & 1.480 [1.450--1.510]\\
Probability/history & 15.121 & 1.519 [1.489--1.549]\\
Seven-component & 17.882 & 1.796 [1.772--1.821]\\
\end{tabular}
\end{ruledtabular}
\end{table}

The uniform comparison measures the benefit of score-based allocation, while the PH comparison measures the additional yield of the full observable combination over the three-component rule. Size-specific results are reported in Appendix Tables~\ref{tab:budget-replay} and \ref{tab:prefix-baseline-size}. The full-score advantage over PH is positive at all five sizes and persists for selection fractions of $10\%$, $25\%$, and $50\%$ and for event thresholds varied by $\pm10\%$ (Appendix~\ref{app:prefix-baseline-sensitivity}). In the additional-depth robustness tests at $n=8,10,12,14$, the unchanged full score gives $1.79$--$2.32$-fold gains over uniform allocation, with every interval above one (Appendix Fig.~\ref{fig:native-depthgrid}c). Upper-to-lower reference-quartile risk ratios and independent-bank susceptibility covariances are also positive relative to their respective null values in every added depth setting (Appendix~\ref{app:native-depthgrid}). These comparisons establish selection value for the fixed structural score across the tested conditions.

\subsection{Structural signatures across sizes and depths}
\label{subsec:structural-results}

\begin{figure*}[t]
\centering
\includegraphics[width=0.96\textwidth]{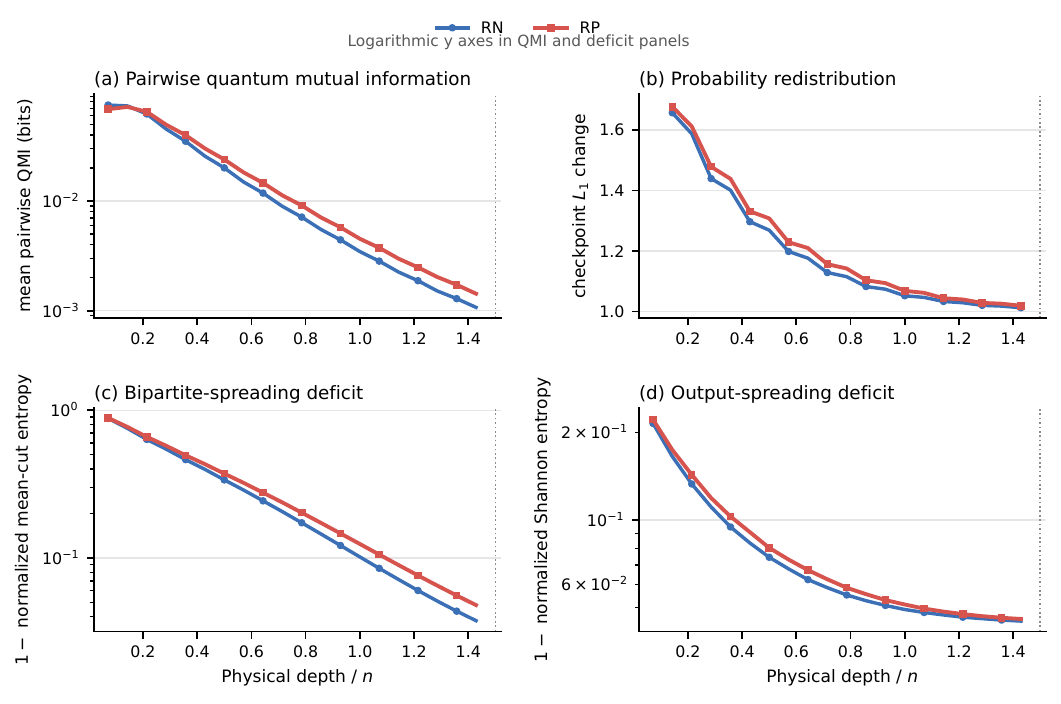}
\caption{\label{fig:trajectories} \textbf{Peak-conditioned circuits separate from their parent ensemble throughout the pre-output trajectory.} This retrospective comparison shows cohort means and circuit-bootstrap $95\%$ intervals for $1000$ RN and $1000$ RP circuits at $n=14$. The horizontal coordinate is physical depth divided by $n$. The dotted line marks the total depth $L/n=1.5$; the terminal checkpoint itself is excluded. Panels show (a) all-pair quantum mutual information, (b) full-distribution $L_1$ redistribution, (c) normalized cut-entanglement deficit, and (d) normalized output-entropy deficit. Panels (a), (c), and (d) use logarithmic axes.}
\end{figure*}

RN and RP separate before the terminal output across every tested size. RP exhibits greater all-pair quantum mutual information (QMI) and stronger layer-to-layer probability redistribution, together with lower normalized cut entanglement and output entropy. The representative $n=14$ trajectories in Fig.~\ref{fig:trajectories} show that these differences persist over broad intervals before the terminal output.

The pre-final effect directions are consistent across all five tested sizes (Appendix Fig.~\ref{fig:cross-size}; Appendix Table~\ref{tab:structural-effects}). Across the twenty endpoint-by-size comparisons, effect magnitudes range from $|g|=0.526$ to $1.531$.

These observables resolve different aspects of the intermediate state. Pairwise QMI measures correlations accessible in two-qubit reduced states, whereas cut entanglement measures entanglement across spatial bipartitions. Their joint behavior shows that stronger pairwise correlations coexist with lower bipartite entanglement in RP. Stronger redistribution adds a dynamical feature: computational-basis probabilities change more between successive layers. A large terminal peak is compatible with nearly maximal Shannon entropy (Appendix~\ref{app:entropy-bounds}), so the peak criterion alone does not require a substantial entropy deficit.

To test whether this separation depends on the prescribed total depth, we generated independent production cohorts at $L_\delta=3n/2-\delta$, with $\delta=1,2,3$, for $n=8,10,12,14$. At the common checkpoint $t=n$, all $48$ new nominal-depth-by-size-by-endpoint contrasts retained their expected directions, and every $95\%$ bootstrap interval excluded zero. Even the weakest contrast remained moderate, $g=0.497$ [$95\%$ CI: $0.410$--$0.585$]. The RN--RP structural separation therefore persists across every implemented setting (Appendix Table~\ref{tab:depth-robustness}).

A trajectory classifier trained only at $n=10,12$ achieves AUROC $0.8043$ [$95\%$ CI: $0.7847$--$0.8233$] when applied without retraining to $1000$ RN and $1000$ RP circuits at $n=16$, demonstrating discrimination beyond the training sizes (Appendix~\ref{app:classifier-specification}).

\subsection{Exact interference underlying realized peaks}
\label{subsec:interference-results}

\begin{figure*}[t]
\centering
\includegraphics[width=0.96\textwidth]{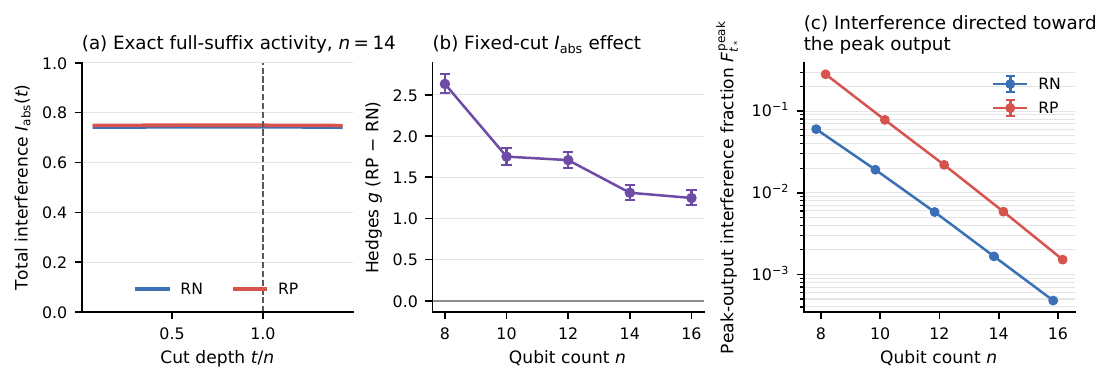}
\caption{\label{fig:interference} \textbf{RP trajectories exhibit greater and more peak-directed quantum interference.} (a) A representative exact full-suffix $\Iabs(t)$ trajectory at $n=14$, shown for visual clarity, with lines denoting cohort means and bands circuit-bootstrap $95\%$ intervals. The dashed line marks the fixed analysis cut $t_*/n=1$. (b) Fixed-cut Hedges effects for $\Iabs$ across $n=8$--$16$ (Table~\ref{tab:interference}). (c) Circuit-level signed interference at the peak output, normalized by the total positive interference mass, $F_{t_*}^{\mathrm{peak}}$, on a logarithmic axis. The peak output is defined retrospectively from each coherent final distribution. Fixed-cut and full-trajectory results use $1000$ circuits per ensemble at every size.}
\end{figure*}

For each circuit, $\Iabs$ is the $L^1$ distance between its coherent output distribution and the distribution obtained by dephasing once in the computational basis at the cut and propagating through the same suffix. It measures the total output-probability redistribution due to interference across that cut. At the fixed analysis cut $t_*=n$, mean $\Iabs$ is larger in RP at every size, with RP-minus-RN Hedges effects ranging from $g=1.25$ to $2.63$ and every $95\%$ interval above zero. The relative increase in mean $\Iabs$ over RN ranges from $0.50\%$ to $13.3\%$ across sizes (Appendix Table~\ref{tab:interference}). Across all five sizes, the full trajectories show that this separation extends over broad pre-output intervals (representative $n=14$ profile in Fig.~\ref{fig:interference}a; complete profiles in the accompanying analysis records).

The signed peak-directed fraction $F_{t_*}^{\mathrm{peak}}$ divides the interference contribution at the realized peak by the total positive interference mass, $\Iabs(t_*)/2$. Computed using the original remaining gates and the location of the observed final peak, this retrospective diagnostic quantifies how interference reinforces that peak. Its numerator concerns one output among $2^n$ possibilities, whereas its denominator sums positive interference over all outputs; its absolute magnitude is therefore interpreted alongside the RN reference at the same size. At the same fixed cut, the mean signed peak-output interference fraction $F_{t_*}^{\mathrm{peak}}$ is larger in RP at all five sizes by factors from $3.17$ to $4.70$ (Fig.~\ref{fig:interference}c), with size-specific $95\%$ circuit-bootstrap intervals reported in Appendix Table~\ref{tab:interference}. RP interference is more strongly directed toward the final peak output throughout the size range, although both ensemble mean fractions decrease with size.

A layer-local analysis finds larger interference in RP even after adjusting for probability redistribution (Appendix Table~\ref{tab:abr}).

At every tested size, the relative increase in the peak-directed fraction is substantially larger than the relative increase in total interference. This contrast identifies the distribution of constructive interference among outputs as a distinguishing feature of RP circuits under their original continuations. The phase-intervention test below then examines whether relative-phase organization also supports peak formation under independently regenerated continuations.

\subsection{Relative-phase organization supports fresh-peak formation}
\label{subsec:phase-results}

The continuation tests establish that the RP advantage survives replacement of its original continuation. We test the functional contribution of relative-phase organization by scrambling phases while preserving every computational-basis population. The phase interaction quantifies the resulting attenuation: the coherent RP-minus-RN event-probability gap minus the corresponding gap after scrambling [Eq.~\eqref{eq:phase-interaction}]. Its magnitude depends on the initial coherent gap as well as the fraction of that gap removed.

\begin{figure*}[t]
\centering
\includegraphics[width=0.96\textwidth]{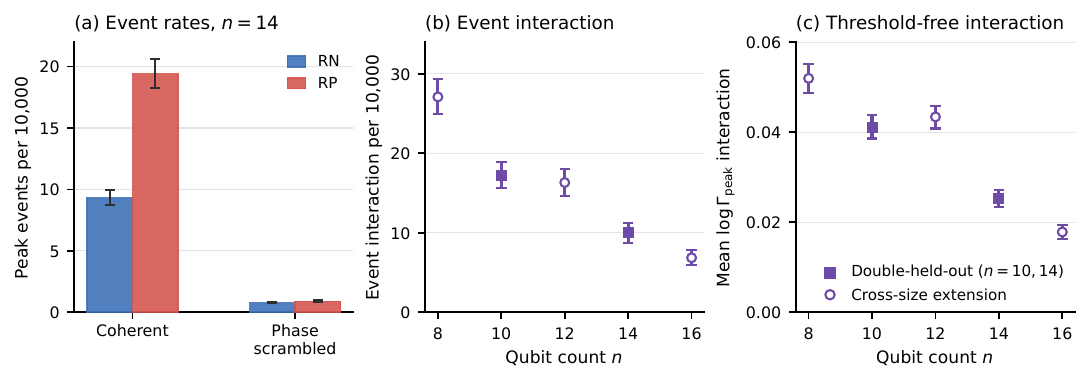}
\caption{\label{fig:phase-intervention} \textbf{The RP fresh-peak advantage depends on its original relative-phase organization across circuit sizes.} (a) Event rates per $10\,000$ for coherent and population-preserving phase-scrambled states at representative size $n=14$. (b) Event interaction, the absolute reduction of the RP--RN event-rate gap, and (c) threshold-free mean-$\log\GammaPeak$ interaction across $n=8$--$16$. Points show difference-in-differences estimates and bars show $95\%$ cluster-bootstrap intervals. Filled squares mark the double-held-out confirmations at $n=10,14$; open circles mark the cross-size extensions. Every interaction interval is positive. Phase scrambling is a controlled pure-state intervention, not computational-basis dephasing.}
\end{figure*}

Population-preserving phase scrambling reduces the coherent RP-minus-RN event-probability gap by $96.3$--$99.4\%$ across all five tested sizes, based on the point estimates. The event interaction is positive throughout this range, spanning $6.848$--$27.098$ events per $10\,000$, with every $95\%$ interval excluding zero (Fig.~\ref{fig:phase-intervention}b and Appendix Table~\ref{tab:phase-crosssize}). Figure~\ref{fig:phase-intervention}a illustrates the effect at the representative size $n=14$: coherent RN and RP states produce $9.329$ and $19.436$ events per $10\,000$, compared with $0.786$ and $0.891$ after phase scrambling.

The threshold-free mean-$\log\GammaPeak$ interaction is likewise positive at every size, ranging from $0.01782$ to $0.05194$, again with all intervals excluding zero (Fig.~\ref{fig:phase-intervention}c and Appendix Table~\ref{tab:phase-crosssize}). The $n=10,14$ confirmations reproduce the initial effect using disjoint states, continuations, and phase seeds; state and bank allocations for every size are listed in Appendix Table~\ref{tab:census}.

The additional-depth robustness tests at $n=8,10,12,14$ reproduce both positive interactions in all $12$ added size--depth combinations, with every interval above zero (Appendix Fig.~\ref{fig:native-depthgrid}b,d). Phase scrambling reduces the coherent RP-minus-RN event-probability gap by $95.2$--$99.7\%$ across these settings, based on the point estimates. This consistency links the selected relative-phase organization to peak formation under fresh continuations across the additional depth settings tested, complementing the primary-depth held-out evidence.

Changing relative phases at fixed basis populations strongly reduces the RP advantage, demonstrating a functional role for the selected phase organization under fresh local continuations. The diagnostic intervention can also change entanglement and reduced-state correlations and produces states outside the original local-circuit preparation law. It therefore probes the organization of the intermediate state as a whole.

A complementary stability test retains RP fresh-event probability and a positive phase interaction after finite prefix-gate deformations at $n=10,12,14$ (Appendix~\ref{app:deformation}).

\subsection{Constructive use of intermediate-state structure}
\label{subsec:construction-results}

To test the constructive value of the observed structure, we compare trajectory-guided and Haar initialization under common refinement, and separately compare multivariate with peak-only intermediate guidance. We construct a starting circuit by selecting newly drawn Haar-gate candidates layer by layer to match an RP reference in observable space. This circuit and a Haar-initialized control then receive the same local-refinement procedure. Reference gates are not used as the candidate bank.

\begin{figure*}[t]
\centering
\includegraphics[width=0.98\textwidth]{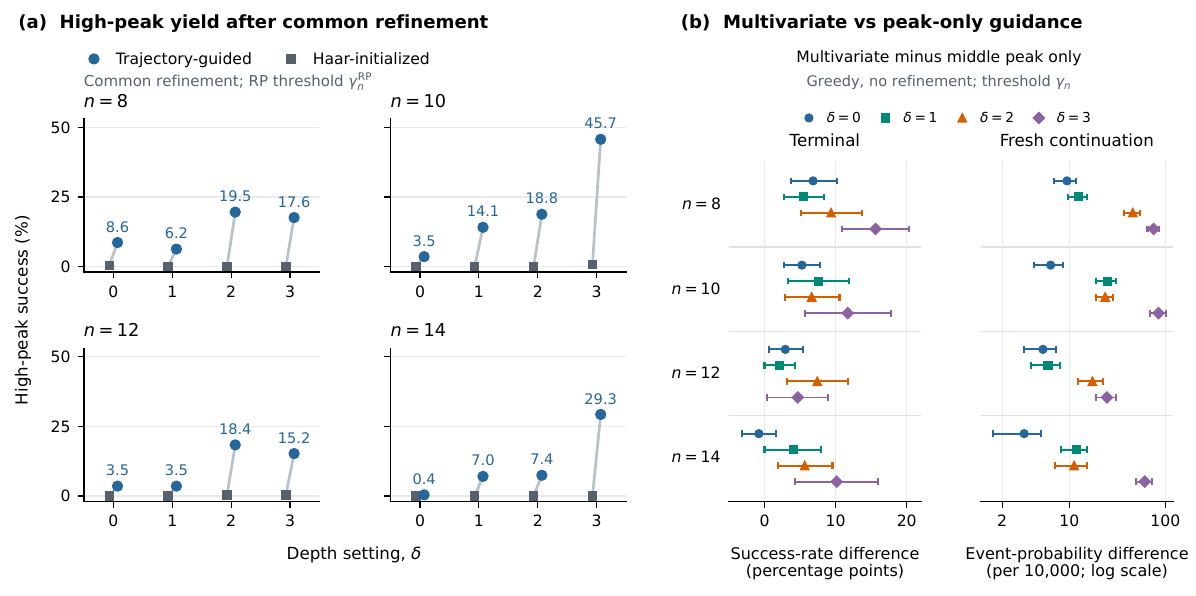}
\caption{\label{fig:construction} \textbf{Structural guidance improves high-peak yield and fresh-peak susceptibility.}
\textbf{(a) High-peak yield after common refinement.} Trajectory-guided initialization (blue circles) is compared with Haar initialization (gray squares), both followed by $1536$-candidate refinement in $256$ paired trials per setting. Success uses the RP thresholds $\gamma_n^{\mathrm{RP}}$; all four size panels share one scale and include every tested $\delta$. Segments connect the two methods within a setting and are not uncertainty bounds. Complete counts and paired $95\%$ intervals are in Appendix Table~\ref{tab:construction}.
\textbf{(b) Multivariate versus peak-only guidance.} The two greedy trackers use identical targets and feature masks at the first and last layers, without lookahead or terminal refinement. Both terminal and fresh events use the calibrated $\gamma_n$, fixed across depths. Points show multivariate minus intermediate peak-only guidance: terminal success on the left (percentage points), and fresh-event probability at cut $t=n$ on the right (per $10\,000$, logarithmic axis). Within each size, symbols run from $\delta=0$ to $3$ top to bottom. The nominal $95\%$ intervals preserve shared-seed pairing after averaging the two fixed-reference contrasts within seed; fresh intervals also resample common suffix columns. All $16$ fresh intervals lie above zero. Complete rates and contrasts are in Appendix Table~\ref{tab:crosssize-ablation}.}
\end{figure*}

At the RP thresholds $\gamma_n^{\mathrm{RP}}$, trajectory-initialized refinement gives a positive high-peak yield difference in all $16$ size--depth settings, with $15$ paired $95\%$ intervals excluding zero (Fig.~\ref{fig:construction}a; Appendix Table~\ref{tab:construction}).

The intermediate-feature ablation connects this constructive use to the structural information. Across the same $16$ size--depth settings, it compares multivariate and peak-only intermediate targets while keeping the endpoint targets fixed, using a greedy tracker without lookahead or terminal refinement. At the calibrated $\gamma_n$, terminal success differences are positive in $15$ settings, with $13$ of the $95\%$ intervals above zero (Fig.~\ref{fig:construction}b; Appendix Table~\ref{tab:crosssize-ablation}).

Under independently regenerated suffixes, multivariate guidance increases peak-event probability in all $16$ settings. The gains range from $3.39$ to $84.85$ events per $10\,000$, with every nominal $95\%$ interval above zero. Their magnitude varies with size and depth, while the additional usefulness of multivariate information persists throughout the tested settings. Thus the constructed intermediate states retain an advantage beyond the particular suffix selected during generation.

The original $n=8$ controls provide complementary feature and ordering tests: the intermediate multivariate contribution remains positive without the redistribution feature, and layer-matched targets improve fresh susceptibility relative to shuffled intermediate targets (Appendix Tables~\ref{tab:construction-ablation} and \ref{tab:construction-ablation-contrasts}). These controls remain separate from the size--depth ablation.

Table~\ref{tab:quantitative-summary} summarizes the main evidence and its inferential roles.

\begin{table*}[t]
\caption{\label{tab:quantitative-summary} Three roles of intermediate-state structure and their tests. Structural and interference diagnostics supply the physical interpretation; estimates and intervals appear with each result.}
\begin{ruledtabular}
\begin{tabular}{lll}
\summarycell{0.17\textwidth}{Role} & \summarycell{0.40\textwidth}{Test} & \summarycell{0.35\textwidth}{What it establishes}\\
\hline
\summarycell{0.17\textwidth}{Prediction} & \summarycell{0.40\textwidth}{Fresh-suffix replacement; unconditional prefix prediction and fixed-baseline allocation (Figs.~\ref{fig:selected-law}, \ref{fig:susceptibility}; Table~\ref{tab:prefix-baselines})} & \summarycell{0.35\textwidth}{Predicts fresh peaks and improves yield over uniform and the tested probability-based rules at equal continuation-evaluation counts.}\\[4pt]
\summarycell{0.17\textwidth}{Phase contribution} & \summarycell{0.40\textwidth}{Population-preserving phase intervention (Fig.~\ref{fig:phase-intervention})} & \summarycell{0.35\textwidth}{Relative-phase organization functionally contributes to the selected-state advantage.}\\[4pt]
\summarycell{0.17\textwidth}{Constructive utility} & \summarycell{0.40\textwidth}{Initialization under common refinement; multivariate versus peak-only guidance (Fig.~\ref{fig:construction})} & \summarycell{0.35\textwidth}{Structural guidance improves high-peak yield and fresh-peak susceptibility in the respective tests.}\\
\end{tabular}
\end{ruledtabular}
\end{table*}

\section{Discussion}
\label{sec:discussion}
\raggedbottom

In the studied regime, finite-depth local dynamics retain input-state sensitivity that global-Haar continuation removes by unitary invariance. Our results show how this sensitivity becomes useful: the frozen prefix score predicts new-peak probability within the unconditional ensemble, while gate-selection comparisons demonstrate the constructive value of RP reference trajectories. Rare-output selection thus identifies structure that remains informative beyond the particular continuation that produced the original peak.

The fixed structural score improves fresh-peak yield over uniform allocation at all five tested sizes, with size-specific yield ratios of $1.55$--$2.08$ at equal continuation-evaluation counts (Appendix Table~\ref{tab:budget-replay}). The additional yield over the probability/history (PH) rule also remains positive throughout the tested range, while decreasing from about $30\%$ at $n=8$ to $6.8\%$ at $n=16$ (Appendix Fig.~\ref{fig:prefix-baseline-sensitivity}a). This trend is observed along a family in which system size and continuation depth increase together, using the same observable types and scoring rule with size-specific reference standardization. The results establish measurable predictive value throughout this family; independently varying size and continuation depth would help resolve what governs the magnitude of the gain.

The structural and interference diagnostics give this predictive relation a physical interpretation. Stronger pairwise correlations and redistribution coexist with lower entanglement and output entropy, and interference is more strongly directed toward the realized peak even where its total activity changes little. Population-preserving phase intervention establishes a functional contribution from relative-phase organization. These distinctions persist across all five tested sizes, even as the peak-directed RP/RN ratio and absolute phase interaction decrease along the size--depth family. Together, these observations connect the organization of the intermediate state to its capacity to form new peaks under local evolution.

Efficient reference generation and observable estimation are priorities for further development. Extending the tests to other connectivity graphs and noise models would probe the scope of the observed behavior. More selective phase interventions could resolve which components of the state organization carry the observed function. Systematic finite-size studies could test whether the observed structural separation reflects a crossover or a transition under varying circuit depth and conditioning strength. Extending the prefix-based analysis to the quantum approximate optimization algorithm (QAOA)~\cite{FarhiGoldstoneGutmann2014} could reveal whether intermediate structure predicts the probability of obtaining high-quality solutions.

\section{Conclusion}
\label{sec:conclusion}

Rare-output conditioning reveals intermediate-state structure that predicts new peak formation under independently regenerated local continuations. Across all five tested sizes, $n=8$--$16$, the frozen structural score improves peak yield over uniform allocation and the tested probability-based rules within an unconditional ensemble. Population-preserving phase intervention establishes a functional contribution from relative-phase organization, while gate-selection comparisons demonstrate the constructive utility of the identified structure.\par

\begin{acknowledgments}
This work was supported by the Quantum Science Center, a National Quantum Information Science Research Center of the U.S. Department of Energy (DOE), operated at Oak Ridge National Laboratory (ORNL).
\end{acknowledgments}

\section*{Author contributions}
S.K. conceived the project. M.K. developed the computational framework, carried out the simulations and analyses, and drafted the manuscript. S.K. and T.H. supervised the research and provided theoretical guidance. All authors contributed to the interpretation of the results, reviewed and revised the manuscript, and approved the final version.




\section*{Accompanying analysis records}
The analysis-records supplement, \nolinkurl{analysis_records.zip}, is provided through the \href{https://1drv.ms/u/c/07C8070870EA6501/AaGHwjPGDzREv_LvJNpha38?e=xAwpw1}{shared research folder}. It contains the frozen score definitions, construction and continuation protocols, statistical notes, verified primary result records, saved continuation-bootstrap draws, paired baseline comparisons, selection-fraction and threshold checks, and numerical summaries accompanying this manuscript. Its file index maps these records to the reported figures and tables.

\onecolumngrid
\raggedbottom
\appendix
\setlength{\intextsep}{6pt}

\section{Circuit ensembles, observables, and estimators}
\label{app:analysis-methods}
\label{sec:methods}

\subsection{Local random architecture and peak amplification}
\label{subsec:random-generator}

We use an open one-dimensional brickwall architecture with independently drawn Haar-$U(4)$ nearest-neighbor gates, generated from complex Gaussian matrices by phase-corrected QR decomposition \cite{Mezzadri2007}.

The primary study and additional-depth robustness tests use implemented, or nominal, layer counts of the form
\begin{equation}
L_\delta=\frac{3n}{2}-\delta,
\qquad \delta\in\{0,1,2,3\}.
\label{eq:depth-robustness}
\end{equation}
The $\delta=0$ case defines the primary ensemble at each of the five system sizes, giving $L=12,15,18,21,$ and $24$ for $n=8,10,12,14,$ and $16$, respectively. This primary setting allows causal influence to span the chain while retaining linear depth and nearest-neighbor connectivity. Unless otherwise noted, main-text figures show this primary setting. To assess robustness to total depth, we add $\delta=1,2,3$ at $n=8,10,12,14$ for structural analysis, continuation replacement, phase intervention, and prefix-score selection (Appendices~\ref{app:depth-robustness} and \ref{app:native-depthgrid}). Given the computational cost of the additional-depth tests, resources at $n=16$ were concentrated on the primary-depth analyses. The separate circuit-construction comparison and intermediate-feature ablation in Fig.~\ref{fig:construction} use all four nominal settings at $n=8,10,12,14$.

All four settings use a depth-$n$ prefix followed by the reverse dagger of an independently drawn forward suffix of depth $\tau_p=n/2-\delta$. We use the outcome-independent boundary $t_*=n$ as the analysis cut; it leaves an implemented continuation depth $\tau_p$, equal to $n/2$ in the primary setting. The Haar-gate distribution is identical on both sides, so the cut marks an analysis horizon rather than a change in the generator or dynamics. When the two layers adjacent to the prefix--suffix boundary act on the same edge matching, their product is itself Haar distributed. Consequently, some nominal-depth settings share the same terminal-unitary distribution, although their fixed-cut state--suffix distributions need not coincide.

To compare circuits at the same prefix depth across different total depths, the four structural observables are evaluated at the common complete-layer checkpoint $t=n$.

For an $n$-qubit state $|\phi\rangle$ and a circuit $U$ acting on $|0^n\rangle$, define
\begin{align}
p_U(z)&=\left|\langle z|U|0^n\rangle\right|^2,\\
\GammaPeak(|\phi\rangle)
&=2^n\max_z\left|\langle z|\phi\rangle\right|^2,\nonumber\\
\GammaPeak(U)
&\equiv\GammaPeak(U|0^n\rangle)
=2^n\max_z p_U(z).
\label{eq:peak-amplification}
\end{align}
Thus $\GammaPeak$ is a functional of the final state that gives its largest computational-basis probability relative to the uniform value $2^{-n}$; $\GammaPeak(U)$ is shorthand for evaluating that functional on $U|0^n\rangle$. At the fixed cut, write the prefix and suffix as $U_{\leq t_*}$ and $U_{>t_*}$, respectively. The intermediate state is
\begin{equation}
|\Psi_{t_*}\rangle=U_{\leq t_*}|0^n\rangle .
\label{eq:analysis-cut-state}
\end{equation}
The suffix induces at the cut the finite-depth local basis
\begin{equation}
|\chi_z\rangle=U_{>t_*}^\dagger|z\rangle .
\label{eq:local-random-basis}
\end{equation}
Under the unconditioned parent law $\mu_{\mathrm{rand}}$, $U_{>t_*}$ is independent of $|\Psi_{t_*}\rangle$. Conditioning on the terminal RP event reweights the retained paths and can correlate a retained prefix with its original suffix. Because $\langle\chi_z|=\langle z|U_{>t_*}$,
\begin{equation}
p_U(z)=\left|\langle\chi_z|\Psi_{t_*}\rangle\right|^2.
\label{eq:random-basis-overlap}
\end{equation}
so $\GammaPeak$ is determined by the largest overlap between the intermediate state and this finite-depth local basis.
For the coherent final distribution, let
\begin{equation}
\zpeak=\arg\max_z p_U(z)
\label{eq:peak-output}
\end{equation}
denote the maximum-probability output string, hereafter called the peak output.

\subsection{RN and RP ensembles}
\label{subsec:ensembles}

Let $\mu_{\mathrm{rand}}$ be the parent circuit law. RN denotes ordinary random reference circuits generated by the same local random-circuit model, while RP circuits are retained for unusually large terminal peaks. Cohort membership rules and source identities are recorded in \nolinkurl{cohort_source_map.json}. The structural cohorts' RP eligibility criterion defines
\begin{equation}
\mu_{\mathrm{RP},n}^{\mathrm{elig}}
=\mu_{\mathrm{rand}}
\left(\,\cdot\mid\GammaPeak(U)\geq\gamma_n^{\mathrm{RP}}\right),
\label{eq:random-laws}
\end{equation}
with
\begin{equation}
\begin{aligned}
\gamma_{8}^{\mathrm{RP}}&=30.0,\quad
\gamma_{10}^{\mathrm{RP}}=31.5,\quad
\gamma_{12}^{\mathrm{RP}}=33.2,\\
\gamma_{14}^{\mathrm{RP}}&=35.1,\quad
\gamma_{16}^{\mathrm{RP}}=36.4.
\end{aligned}
\label{eq:rp-eligibility}
\end{equation}
Here $\mu_{\mathrm{RP},n}^{\mathrm{elig}}$ denotes the threshold-eligible population; the analyzed RP cohorts are the frozen subsets described below. The size dependence follows the extreme-value scale: for a Haar state in dimension $D=2^n$, $\E[\GammaPeak]=H_D\simeq n\ln2+\gamma_{\mathrm E}$.

At the primary depth, the structural analysis uses $1000$ RN and $1000$ RP circuits at each of the five sizes, $n=8$--$16$. At $n=8$--$14$, each primary structural RP cohort contains the $1000$ largest-$\GammaPeak$ eligible circuits from its frozen production pool. No structural or interference observable entered selection. The additional-depth structural cohorts ($\delta=1,2,3$) at $n=8,10,12,14$ also contain $1000$ RN and $1000$ RP circuits per setting. They were generated independently under the same local gate law and frozen ensemble definitions, retaining the primary size-specific thresholds $\gamma_n^{\mathrm{RP}}$. These cohorts test persistence across the prescribed nominal depths (Appendix Table~\ref{tab:depth-robustness}); their RP production pools also supply reference trajectories for the construction study (Appendix~\ref{app:construction}).

At the primary $n=16$ depth, the structural, frozen trajectory-classifier, fixed-cut and full-trajectory exact-interference, layer-local decomposition, and phase analyses use the same $1000$ RN and $1000$ RP circuits. Table~\ref{tab:dataset} summarizes the primary structural cohorts.

\begin{table*}[t]
\caption{\label{tab:dataset} Primary RN--RP structural dataset. RP peak statistics are median [interquartile range] values of $\GammaPeak$ in the retained cohorts.}
\begin{ruledtabular}
\begin{tabular}{cccccc}
$n$ & Nominal depth & RN count & RP count & RP threshold & RP $\GammaPeak$ median [IQR]\\
\hline
8  & 12 & 1000 & 1000 & 30.0 & 31.81 [30.88--33.13]\\
10 & 15 & 1000 & 1000 & 31.5 & 33.38 [32.53--35.08]\\
12 & 18 & 1000 & 1000 & 33.2 & 34.86 [33.90--36.42]\\
14 & 21 & 1000 & 1000 & 35.1 & 36.41 [35.62--37.90]\\
16 & 24 & 1000 & 1000 & 36.4 & 37.55 [36.84--38.74]\\
\end{tabular}
\end{ruledtabular}
\end{table*}

For a complex Haar state, the basis probabilities follow $\operatorname{Dirichlet}(1,\ldots,1)$ \cite{ZyczkowskiSommers2001}. If $M=\max_i p_i$, uniform-simplex geometry gives
\begin{equation}
\Prob(M\leq m)
=
\sum_{k=0}^{\min(D,\lfloor1/m\rfloor)}
(-1)^k{D\choose k}(1-km)^{D-1}.
\label{eq:haar-maximum}
\end{equation}
At $n=16$, this reference gives $\Prob(\GammaPeak\geq36.4)\simeq1.0093\times10^{-11}$. The implemented local architecture produced the same event only about once per $34.2$ million circuit draws. The global-Haar value is only a diagnostic reference and is not the circuit generator.

\subsection{Trajectory observables}
\label{subsec:observables}

Observables are evaluated after every complete physical brickwall layer. Serialized single-gate checkpoints are excluded, as is the terminal checkpoint that defines cohort membership. For reduced states $\rho_i,\rho_j,\rho_{ij}$, quantum mutual information and its all-pair mean are
\begin{align}
I_Q(i:j)&=S(\rho_i)+S(\rho_j)-S(\rho_{ij}),\\
S(\rho)&=-\operatorname{Tr}(\rho\log_2\rho),\\
\overline I_Q&=\binom{n}{2}^{-1}\sum_{i<j}I_Q(i:j).
\label{eq:qmi}
\end{align}
For a contiguous cut $c$, normalized cut entanglement and its cut average are
\begin{equation}
E_c=\frac{S(\rho_{1:c})}{\min(c,n-c)},
\qquad
\overline E_{\mathrm{cut}}=\frac{1}{n-1}\sum_{c=1}^{n-1}E_c.
\label{eq:cut-entanglement}
\end{equation}
For computational-basis probabilities $p(z)$, normalized Shannon entropy is
\begin{equation}
\widetilde H_1(p)
=
-\frac1n\sum_zp(z)\log_2p(z).
\label{eq:shannon}
\end{equation}
In this subsection, $p_\ell$ denotes the computational-basis distribution at complete-layer checkpoint $\ell$. Full-distribution redistribution between consecutive complete layers is
\begin{equation}
R_\ell=\sum_z|p_\ell(z)-p_{\ell-1}(z)|,
\qquad \ell\geq2.
\label{eq:redistribution}
\end{equation}
No factor $1/2$ is included. The first checkpoint has no redistribution value.

For the display in Fig.~\ref{fig:trajectories}, entanglement and output spreading are shown as deficits
\begin{equation}
D_E=1-\overline E_{\mathrm{cut}},
\qquad
D_H=1-\widetilde H_1.
\label{eq:deficits}
\end{equation}
The transformation and logarithmic axes are display choices. Effect sizes use the original observables.

\subsection{Exact full-suffix interference}
\label{subsec:interference-method}

Let the physical layers be ordered as $U=U_L\cdots U_1$. At a cut $t$ immediately after complete physical layer $t$, write
\begin{equation}
\begin{aligned}
|\Psi_t\rangle
&=U_t\cdots U_1|0^n\rangle\\
&=\sum_x a_t(x)|x\rangle,\\
U_{>t}&=U_L\cdots U_{t+1}.
\end{aligned}
\end{equation}
The coherent output probability decomposes exactly as
\begin{align}
p_U(z)
&=\left|\sum_x\left(U_{>t}\right)_{zx}a_t(x)\right|^2
=p_{\mathrm{inc}}^{(t)}(z)+I^{(t)}(z),\\
p_{\mathrm{inc}}^{(t)}(z)
&=\sum_x\left|\left(U_{>t}\right)_{zx}\right|^2|a_t(x)|^2,\\
I^{(t)}(z)
&=2\operatorname{Re}\sum_{x<y}
\left(U_{>t}\right)_{zx}\left(U_{>t}\right)_{zy}^{*}a_t(x)a_t^*(y).
\label{eq:interference-decomposition}
\end{align}
Here $p_{\mathrm{inc}}^{(t)}$ is the cut-dependent output distribution after computational-basis dephasing exactly once at the cut, followed by fully coherent propagation through the original suffix. The coherent final distribution $p_U$ is independent of the choice of exact cut; only its decomposition changes with $t$. The signed interference contribution is
\begin{equation}
I^{(t)}(z)=p_U(z)-p_{\mathrm{inc}}^{(t)}(z),
\qquad
\sum_z I^{(t)}(z)=0.
\label{eq:signed-interference}
\end{equation}
We quantify total interference activity by
\begin{equation}
\Iabs(t)=\sum_z|I^{(t)}(z)|.
\label{eq:iabs}
\end{equation}
Therefore $\Iabs(t)/2$ equals both the total positive interference mass and the total-variation distance between $p_U$ and $p_{\mathrm{inc}}^{(t)}$.

Using the peak output defined in Eq.~(\ref{eq:peak-output}), the signed peak-output interference fraction is
\begin{equation}
F_t^{\mathrm{peak}}
=\frac{I^{(t)}(\zpeak)}{\Iabs(t)/2}.
\label{eq:peak-output-fraction}
\end{equation}
We use ``peak-directed interference fraction'' as a short description. Positive $F_t^{\mathrm{peak}}$ means that interference constructively reinforces the final peak output. Because $\zpeak$ is defined retrospectively from the coherent final distribution, $F_t^{\mathrm{peak}}$ is an exact decomposition of a realized peak, not a prospective predictor or a causal intervention. Initial and terminal cuts, for which $\Iabs=0$, are excluded. Each circuit-level fraction is formed before ensemble averaging; the reported RP/RN contrast is the ratio of the resulting ensemble means.

The calculation uses exact complete enumeration of all computational-basis sources. No Monte Carlo sampling or tensor-network truncation enters $\Iabs$. Both the fixed-cut and full-trajectory analyses use $1000$ RN and $1000$ RP circuits at each tested size in $n=8$--$16$, totaling $10\,000$ circuits. Full trajectories include every nontrivial complete-layer cut. Numerical cross-checks and their tolerances are documented in the accompanying analysis records.

\subsection{Fresh-continuation susceptibility and prospective prefix scoring}
\label{subsec:continuations}

Let $t_*=n$ be the fixed outcome-independent analysis cut defined above. Fresh-continuation events use size-specific thresholds $\gamma_n$ calibrated as the empirical $0.999$ quantile of $10^7$ independent state--continuation pairs at the primary depth. Table~\ref{tab:threshold-roles} reports rounded values; calculations use unrounded thresholds. Calibration does not use ensemble labels or prefix scores. The depth extensions retain these absolute thresholds rather than recalibrating each depth to the same tail probability. We write
\begin{equation}
q_{n,t_*}(\Psi_{t_*})
=
\Prob_V\!\left[
\GammaPeak(V|\Psi_{t_*}\rangle)\geq\gamma_n
\mid\Psi_{t_*}
\right].
\label{eq:empirical-susceptibility}
\end{equation}

Each fresh suffix $V$ is drawn independently of the retained states and their original retention events. These experiments therefore restore prospective prefix--suffix independence, even though such independence need not hold between an RP state and its original suffix.

The unconditional prediction experiment uses $2048$ states per size for $n=8,10,12,14$ and $1000$ states at $n=16$. At every size, two shared-column banks each contain $16\,384$ suffixes applied to every prefix. A seven-component prefix score uses only information available at or before $t_*$. In fixed order, its components are all-pair QMI, adjacent-pair QMI, redistribution at the last prefix transition, normalized Shannon entropy, collision ratio, normalized mean cut entanglement, and mean leading Schmidt weight. The components not defined above are
\begin{align}
I_Q^{\mathrm{adj}}
&=\frac{1}{n-1}\sum_{a=1}^{n-1}I_Q(a:a+1),\\
C(p)&=2^n\sum_z p(z)^2,\\
w_c&=\max_k
\frac{\varsigma_{c,k}^2}{\sum_\ell\varsigma_{c,\ell}^2},
\qquad
x_{\mathrm{Sch}}=\frac{1}{n-1}\sum_{c=1}^{n-1}w_c,
\label{eq:prefix-components}
\end{align}
where $p$ is the computational-basis distribution at the analysis cut and $\varsigma_{c,k}$ are the Schmidt coefficients across spatial cut $c$. The feature orientations are informed by the RP-associated structural pattern and frozen before fresh-outcome evaluation. Each component has a fixed RP-associated orientation and is standardized using an independent $10\,000$-state reference bank. Their equal-weight mean is reference-standardized to give $s_{\mathrm{pre},ni}$; neither terminal nor fresh-continuation outcomes of the evaluated circuit are used to compute or retune this score. Appendix~\ref{app:score-statistics} gives the exact standardization and reference records. This prefix-only score is distinct from the trajectory classifier in Sec.~\ref{subsec:cross-size-classification}.

At $n=8,10,12,14$, two additional row-specific banks each contain $16\,384$ independently drawn suffixes per prefix and test repeatability through the covariance of their estimated state hit rates. Independent continuation-sampling noise has zero cross-covariance, so this statistic measures repeatable between-state susceptibility variation; its estimator is specified in Appendix~\ref{app:score-statistics}.

At each of the five sizes, the continuation-replacement experiment discards the original continuations of $1000$ RN states and $1000$ RP states. Both cohorts receive two common fresh banks of $16\,384$ continuations each, with continuation depth $n/2$ at the primary setting. State and continuation clustering are preserved in inference.

The frozen-budget replay uses the already generated unconditional continuation banks and the prespecified prefix score. It compares allocating a fixed number of continuation evaluations to the reference-defined upper score quartile with allocating the same number uniformly over all prefixes. Appendix~\ref{app:baseline-policy-comparison} defines the matched-count probability-based baselines and paired comparisons. The across-size summary assigns equal weight to $n=8$--$16$.

At $n=8$--$14$, continuation replacement and prediction use independently generated state cohorts and suffix banks; at $n=16$, prediction and policy replay use the RN states and banks from the replacement experiment. Absolute baseline rates are experiment-specific estimates, and effects are defined within the corresponding bank design. Reported continuation intervals condition on the calibrated thresholds; they do not propagate calibration-sample uncertainty.

In the additional-depth robustness tests, we repeat continuation replacement and score-based allocation at $n=8,10,12,14$ for $\delta=1,2,3$, using the depth-matched RN/RP cohorts and the unchanged unconditional prefixes and prefix score. Fresh suffixes follow the corresponding remaining-depth law, while the cut and absolute event thresholds remain fixed. The bank design and complete depth-specific results are given in Appendix~\ref{app:native-depthgrid}.

\subsection{Population-preserving phase intervention}
\label{subsec:phase-intervention}

The phase intervention tests whether the original relative-phase organization contributes to the RN--RP fresh-peak difference. For
\begin{equation}
a_{t_*}(x)=|a_{t_*}(x)|e^{i\phi_x},
\end{equation}
we form eight pure-state replicas
\begin{equation}
a_{t_*}^{(k)}(x)=|a_{t_*}(x)|e^{i\theta_x^{(k)}},
\qquad
\theta_x^{(k)}\overset{\mathrm{iid}}{\sim}\operatorname{Unif}[0,2\pi).
\label{eq:phase-scrambling}
\end{equation}
This intervention preserves every computational-basis population exactly while replacing the relative phases. It is phase scrambling, not dephasing. In addition to the primary-depth tests below, the same intervention is evaluated on every depth-extension RN/RP cohort using fresh suffixes and phase draws (Appendix~\ref{app:native-depthgrid}).

The phase intervention is evaluated at all five system sizes. At $n=10,14$, double-held-out confirmations use states, suffix columns, and phase seeds disjoint from the initial experiment; the other sizes provide cross-size extensions. At $n=16$, the test uses the first $2048$ fixed columns from each existing A/B continuation bank, reuses their coherent outcomes, and applies eight fresh phase replicas per state, paired across banks. The threshold $\gamma_{16}$ remains fixed. State allocations and index partitions are detailed in Appendix~\ref{app:census}. The primary interaction is
\begin{equation}
\Delta_n^{\mathrm{phase}}
=
(q_{\mathrm{RP}}^{\mathrm{coh}}-q_{\mathrm{RN}}^{\mathrm{coh}})
-
(q_{\mathrm{RP}}^{\mathrm{scr}}-q_{\mathrm{RN}}^{\mathrm{scr}}).
\label{eq:phase-interaction}
\end{equation}
Here $q_{\mathcal E}^{\mathrm{coh}}$ and $q_{\mathcal E}^{\mathrm{scr}}$ denote ensemble-averaged fresh-continuation event probabilities at fixed $n$, for $\mathcal E\in\{\mathrm{RN},\mathrm{RP}\}$; the scrambled probability additionally averages over the randomized phases of the pure-state replicas. The same difference-in-differences is evaluated for mean $\log\GammaPeak$ as a threshold-free confirmation. All logarithms of $\GammaPeak$ in the phase-intervention analysis are natural logarithms.

\subsection{Cross-size trajectory classification}
\label{subsec:cross-size-classification}

The cross-size stress test uses a separate $150$-component trajectory representation obtained by sampling $15$ pre-final structural curves at ten normalized depths. An elastic-net score is fitted using only $n=10,12$ RN and RP circuits, with terminal $\GammaPeak$, peak-output identity, and terminal-checkpoint features excluded. The fitted preprocessing and coefficients are then applied without refitting, score reversal, or threshold recalibration to the independent $n=16$ cohort. The representation includes pre-final checkpoints after $t_*$ and therefore evaluates trajectory classification rather than prefix-only prediction. This score tests RN--RP separation at a larger system size and is distinct from the seven-component prefix score used to predict peaks under fresh continuations. Appendix~\ref{app:classifier-specification} specifies the representation, training procedure, and accompanying model record.

\subsection{Trajectory-guided circuit construction}
\label{subsec:construction-method}

At $n=8,10,12,14$ and each $L_\delta$, we use a size- and depth-matched library of $256$ RP observable trajectories to guide new gate selection. A reference $r$ is selected without screening the generated outcome and kept fixed throughout each trial. For layer candidate $j$, define
\begin{equation}
\mathcal L_\ell^{(r)}(j)
=\frac{1}{|\mathcal F|}\sum_{k\in\mathcal F}
\left(\frac{f_{\ell k}^{(j)}-f_{\ell k}^{(r)}}{s_{\ell k}}\right)^2.
\label{eq:trajectory-matching-loss}
\end{equation}
Here $\mathcal F$ contains seven observables and $s_{\ell k}$ is a fixed RN-reference scale. The features, reference sources, and one-layer lookahead rule are specified in Appendix~\ref{app:construction}. The classifier ranks pre-final trajectories, the prefix score predicts fresh outcomes, and this objective selects new gates. Gate candidates are Haar drawn, but their trajectory-guided selection changes the resulting circuit law. Constructed circuits are not added to the RN or RP analysis cohorts.

For each setting, we compare trajectory and Haar initializations in $M=256$ paired trials. Both receive the same $1536$-candidate adaptive local refinement, specified in Appendix~\ref{app:construction}. All trials are included regardless of success. Construction success is evaluated using the RP thresholds $\gamma_n^{\mathrm{RP}}$ defined in Eq.~\eqref{eq:rp-eligibility}. Writing $Y_i^a=\mathbf 1\{\GammaPeak(U_i^a)\geq\gamma_n^{\mathrm{RP}}\}$ for the final output of arm $a$, the paired yield difference is
\begin{equation}
\widehat{\Delta P}_{n,\delta}
=\frac1M\sum_{i=1}^{M}
\left(Y_i^{\mathrm{T+R}}-Y_i^{\mathrm{H+R}}\right),
\label{eq:construction-gain}
\end{equation}
where T+R and H+R denote trajectory- and Haar-initialized refinement. Paired-trial bootstrap intervals and target-cluster sensitivity preserve the reference reuse. The matched budget covers the $1536$ refinement candidates; trajectory initialization and reference generation are outside this budget.

A separate feature ablation at the same four sizes and all four nominal depth settings compares multivariate and peak-only intermediate targets while keeping first- and last-layer targets and feature masks identical. Both arms use greedy selection without lookahead or terminal refinement. Terminal success is the primary endpoint; susceptibility at cut $t=n$ under independently generated suffixes is the secondary endpoint. Both use the calibrated thresholds $\gamma_n$ defined in Sec.~\ref{subsec:continuations}, held fixed across depths. Sample allocations, paired inference, and the original $n=8$ controls are reported in Appendix~\ref{app:construction}.

\subsection{Statistical analysis}
\label{subsec:statistics}

The circuit or prefix state is the primary state-level inference unit. With $L_n=3n/2$ complete layers, the pre-final scalar for trajectory observable $M$ is
\begin{equation}
S_{i,M}^{\mathrm{pre}}
=\frac{1}{L_n-1}\sum_{\ell=1}^{L_n-1}M_i(\ell),
\qquad
M\in\{\overline I_Q,\overline E_{\mathrm{cut}},\widetilde H_1\}.
\label{eq:prefinal-scalar}
\end{equation}
For redistribution, which is defined between successive checkpoints, we use
\begin{equation}
S_{i,R}^{\mathrm{pre}}
=\frac{1}{L_n-2}\sum_{\ell=2}^{L_n-1}R_{i,\ell}.
\label{eq:prefinal-redistribution}
\end{equation}
Thus ``pre-final'' denotes a circuit-level trajectory mean rather than a single penultimate checkpoint; the terminal checkpoint is excluded, and redistribution also excludes the first checkpoint. RN--RP contrasts use bias-corrected Hedges $g$, with positive values denoting larger RP means \cite{Hedges1981}.

Structural and exact-interference intervals resample circuits independently within each ensemble. Continuation inference resamples state and suffix axes while preserving shared-bank pairing; the primary phase-intervention inference additionally resamples phase replicas within state. The estimators, replicate counts, and ratio intervals are specified in Appendix~\ref{app:score-statistics}, following standard nonparametric practice \cite{EfronTibshirani1993}. Appendix~\ref{app:native-depthgrid} specifies the two-way bootstrap for the depth extensions, in which phase replicas are averaged within state.

\section{Sharp entropy bounds at fixed maximum}
\label{app:entropy-bounds}

For a $D$-outcome probability vector with exact maximum $m$, let $k=\lfloor1/m\rfloor$ and $r=1-km$. Here $H_1(p)=-\sum_i p_i\log_2p_i$ is the unnormalized Shannon entropy in bits. For $D=2^n$, it is related to the normalized entropy used in the main text by $\widetilde H_1(p)=H_1(p)/n$. Majorization and Schur concavity give
\begin{align}
-km\log_2m-r\log_2r
&\leq H_1 \nonumber\\
&\leq h_2(m)+(1-m)\log_2(D-1),
\label{eq:shannon-bounds}\\
m^2+\frac{(1-m)^2}{D-1}
&\leq \sum_i p_i^2 \leq km^2+r^2.
\label{eq:collision-bounds}
\end{align}
At $n=14$, the minimum Shannon-entropy deficits enforced by terminal maxima $\GammaPeak=30$ and $35.1$ are $0.00643$ and $0.00800$ bit, obtained when the remaining probability mass is distributed as uniformly as allowed. A large terminal peak is compatible with a nearly maximal Shannon entropy; the fixed-maximum bound therefore does not require a large global entropy reduction.

\section{Structural robustness across sizes and depths}
\label{app:depth-robustness}

Table~\ref{tab:structural-effects} reports RP-minus-RN Hedges $g$ for circuit-level pre-final trajectory means at the primary depth across all five sizes. Table~\ref{tab:depth-robustness} assesses depth robustness at $n=8,10,12,14$, reporting the corresponding effects at the common checkpoint $t=n$ for the primary and three additional nominal depths.

\begin{table}[H]
\caption{\label{tab:structural-effects} Pre-final trajectory-mean effects, RP minus RN, corresponding to Fig.~\ref{fig:cross-size}. Entries are Hedges $g$. Each size uses $1000$ RN and $1000$ RP circuits.}
\begin{ruledtabular}
\begin{tabular}{crrrr}
$n$ & All-pair QMI & Redistribution & Normalized cut entanglement & Normalized entropy\\
\hline
8  & 1.322 & 1.006 & -1.531 & -0.996\\
10 & 0.990 & 0.840 & -1.193 & -0.792\\
12 & 1.056 & 0.932 & -1.119 & -0.719\\
14 & 0.873 & 0.780 & -0.951 & -0.588\\
16 & 0.778 & 0.769 & -0.898 & -0.526\\
\end{tabular}
\end{ruledtabular}
\end{table}

\begin{figure}[H]
\centering
\includegraphics[width=0.82\textwidth]{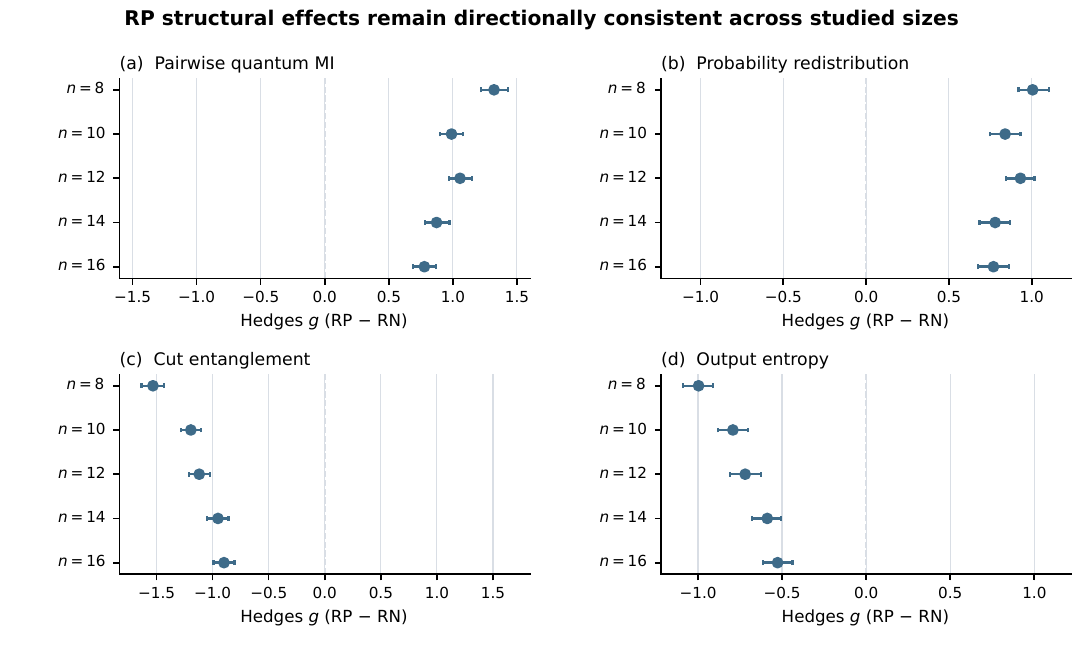}
\caption{\label{fig:cross-size} \textbf{The RN--RP structural signature persists across the tested sizes.} Hedges-$g$ effects and $95\%$ ensemble-stratified circuit-bootstrap intervals are shown for the four trajectory endpoints. Each of the five sizes uses $1000$ RN and $1000$ RP circuits. Effects express the RP--RN separation in units of the pooled within-size standard deviation. Their directions persist, while their magnitudes vary with $n$.}
\end{figure}

\begin{table}[H]
\caption{\label{tab:depth-robustness} Depth robustness of fixed-checkpoint structural effects at $n=8,10,12,14$. Entries are RP-minus-RN Hedges $g$ with $95\%$ ensemble-stratified circuit-bootstrap intervals at the common completed-layer checkpoint $t=n$. The $L=3n/2$ rows use the primary cohorts; each $L-1,L-2,$ and $L-3$ row uses an independently generated production cohort of $1000$ RN and $1000$ RP circuits. Nominal-depth equivalence is discussed in Sec.~\ref{subsec:random-generator}. Positive effects are expected for all-pair QMI and redistribution, and negative effects for normalized cut entanglement and output entropy.}
\squeezetable
\begin{ruledtabular}
\begin{tabular}{ccrrrr}
Nominal depth & $n$ & All-pair QMI & Redistribution & Normalized cut entanglement & Normalized entropy\\
\hline
$L$ & 8 & 1.432 [1.334, 1.533] & 0.501 [0.408, 0.591] & -1.640 [-1.744, -1.540] & -0.735 [-0.819, -0.653]\\
$L$ & 10 & 1.176 [1.083, 1.268] & 0.625 [0.536, 0.714] & -1.347 [-1.444, -1.254] & -0.713 [-0.801, -0.625]\\
$L$ & 12 & 1.180 [1.090, 1.276] & 0.733 [0.649, 0.820] & -1.285 [-1.379, -1.192] & -0.827 [-0.915, -0.740]\\
$L$ & 14 & 0.995 [0.911, 1.083] & 0.672 [0.583, 0.765] & -1.114 [-1.210, -1.023] & -0.738 [-0.830, -0.653]\\
$L-1$ & 8 & 1.440 [1.346, 1.536] & 0.497 [0.410, 0.585] & -1.654 [-1.764, -1.551] & -0.795 [-0.879, -0.712]\\
$L-1$ & 10 & 1.161 [1.074, 1.256] & 0.629 [0.545, 0.718] & -1.250 [-1.346, -1.153] & -0.696 [-0.782, -0.610]\\
$L-1$ & 12 & 1.116 [1.023, 1.208] & 0.623 [0.537, 0.711] & -1.251 [-1.349, -1.155] & -0.697 [-0.788, -0.612]\\
$L-1$ & 14 & 1.083 [0.988, 1.175] & 0.759 [0.671, 0.848] & -1.168 [-1.266, -1.079] & -0.875 [-0.967, -0.781]\\
$L-2$ & 8 & 1.489 [1.394, 1.597] & 0.615 [0.529, 0.703] & -1.691 [-1.799, -1.589] & -0.823 [-0.912, -0.738]\\
$L-2$ & 10 & 1.313 [1.226, 1.405] & 0.591 [0.503, 0.684] & -1.407 [-1.503, -1.315] & -0.719 [-0.805, -0.635]\\
$L-2$ & 12 & 1.115 [1.023, 1.207] & 0.590 [0.500, 0.681] & -1.182 [-1.279, -1.087] & -0.709 [-0.799, -0.620]\\
$L-2$ & 14 & 1.118 [1.032, 1.206] & 0.803 [0.718, 0.891] & -1.260 [-1.356, -1.168] & -0.786 [-0.879, -0.698]\\
$L-3$ & 8 & 1.538 [1.438, 1.643] & 0.527 [0.439, 0.616] & -1.870 [-1.978, -1.770] & -0.863 [-0.948, -0.778]\\
$L-3$ & 10 & 1.335 [1.243, 1.432] & 0.615 [0.531, 0.705] & -1.422 [-1.520, -1.332] & -0.843 [-0.927, -0.760]\\
$L-3$ & 12 & 1.206 [1.116, 1.303] & 0.614 [0.523, 0.704] & -1.324 [-1.421, -1.232] & -0.736 [-0.826, -0.651]\\
$L-3$ & 14 & 1.181 [1.091, 1.276] & 0.812 [0.726, 0.908] & -1.305 [-1.402, -1.211] & -0.912 [-1.003, -0.824]\\
\end{tabular}
\end{ruledtabular}
\end{table}

\section{Exact-interference results}
\label{app:interference-table}

For the signed peak-output fraction, standardized effects range from $g=11.801$ to $15.005$ across all five tested sizes. Table~\ref{tab:interference} reports the mean fractions and their RP/RN ratios.

\begin{table}[H]
\caption{\label{tab:interference} Exact interference results at the fixed analysis cut $t_*=n$, using $1000$ RN and $1000$ RP circuits at every size. Intervals are $10\,000$-replicate independent circuit-bootstrap percentile intervals.}
\begin{ruledtabular}
\begin{tabular}{crrrrrr}
$n$ & RN mean $\Iabs$ & RP mean $\Iabs$ & $\Iabs$ Hedges $g$ [95\% CI] & RN mean $F_{t_*}^{\mathrm{peak}}$ & RP mean $F_{t_*}^{\mathrm{peak}}$ & RP/RN ratio [95\% CI]\\
\hline
8  & 0.750619 & 0.850258 & 2.632 [2.520--2.752] & 0.060163 & 0.282760 & 4.70 [4.62--4.78]\\
10 & 0.747540 & 0.780737 & 1.750 [1.650--1.854] & 0.019157 & 0.078009 & 4.07 [4.02--4.13]\\
12 & 0.748015 & 0.766243 & 1.708 [1.614--1.810] & 0.005810 & 0.021967 & 3.78 [3.74--3.83]\\
14 & 0.742671 & 0.749447 & 1.312 [1.221--1.407] & 0.001665 & 0.005869 & 3.53 [3.49--3.56]\\
16 & 0.741604 & 0.745342 & 1.249 [1.159--1.341] & 0.000479 & 0.001518 & 3.17 [3.14--3.20]\\
\end{tabular}
\end{ruledtabular}
\end{table}

\section{Cross-size phase-intervention results}
\label{app:phase-crosssize}

\begin{table}[H]
\caption{\label{tab:phase-crosssize} Phase-intervention results across circuit sizes. The four event-rate columns and the event interaction are reported per $10\,000$ fresh continuations. Both interactions use the contrast $(\mathrm{RP}-\mathrm{RN})_{\mathrm{coh}}-(\mathrm{RP}-\mathrm{RN})_{\mathrm{scr}}$. The final column reports the corresponding interaction in mean $\log\GammaPeak$ and is not rescaled. ``Coh.'' denotes coherent states and ``scr.'' denotes population-preserving phase-scrambled states. Intervals are $95\%$ additive state--suffix cluster-bootstrap intervals with phase replicas resampled within state. All event thresholds use the size-specific empirical $0.999$ calibration in Table~\ref{tab:threshold-roles}.}
\begin{ruledtabular}
\begin{tabular}{crrrrrr}
$n$ & RN coh. & RP coh. & RN scr. & RP scr. & $10^4\Delta_n^{\mathrm{phase}}$ [95\% CI] & Mean-$\log\GammaPeak$ interaction [95\% CI]\\
\hline
8  & 8.476 & 36.615 & 0.612 & 1.652 & 27.098 [24.891--29.341] & 0.05194 [0.04875--0.05510]\\
10 & 8.338 & 25.832 & 0.661 & 0.959 & 17.195 [15.575--18.862] & 0.04113 [0.03858--0.04373]\\
12 & 7.616 & 24.033 & 0.231 & 0.333 & 16.315 [14.653--18.001] & 0.04338 [0.04080--0.04588]\\
14 & 9.329 & 19.436 & 0.786 & 0.891 & 10.002 [8.719--11.277] & 0.02523 [0.02334--0.02716]\\
16 & 10.000 & 16.899 & 0.791 & 0.842 & 6.848 [5.884--7.835] & 0.01782 [0.01630--0.01927]\\
\end{tabular}
\end{ruledtabular}
\end{table}

\section{Interference and redistribution decomposition}
\label{app:abr}

For layer $\ell$, let $p_{\ell-1}$ and $p_\ell$ be the coherent distributions before and after the layer, and let $p_\ell^{\mathrm{inc}}$ be the population-only one-layer counterfactual obtained by propagating $p_{\ell-1}$ through the squared gate moduli. Define $i_\ell=p_\ell-p_\ell^{\mathrm{inc}}$, $b_\ell=p_\ell^{\mathrm{inc}}-p_{\ell-1}$, and $r_\ell=p_\ell-p_{\ell-1}=b_\ell+i_\ell$. The layer-local components are $A_\ell=\|i_\ell\|_1$, $B_\ell=\|b_\ell\|_1$, $R_\ell=\|r_\ell\|_1$, and $C_\ell=A_\ell+B_\ell-R_\ell$; Table~\ref{tab:abr} reports effects for their pre-final circuit means. Every component is larger in RP. Regressing the circuit-mean $A$ on the circuit-mean $R$ leaves a positive RP-minus-RN residual effect at every size, indicating that interference activity is not a relabeling of redistribution alone. The RP pointwise mean $A_\ell$ exceeds RN at every nontrivial layer across $n=8$--$16$, placing the separation throughout the trajectory rather than only at the fixed analysis cut.

\begin{table}[H]
\caption{\label{tab:abr} Pre-final Hedges effects for the layer-local $A$--$B$--$R$ decomposition. Every size uses $1000$ RN and $1000$ RP circuits.}
\begin{ruledtabular}
\begin{tabular}{crrrrr}
$n$ & $A$ & $B$ & $R$ & $C$ & Residual $A$ after $A\sim R$\\
\hline
8  & 1.072 & 0.920 & 1.006 & 0.715 & 0.473\\
10 & 0.902 & 0.808 & 0.840 & 0.757 & 0.369\\
12 & 0.942 & 0.860 & 0.932 & 0.770 & 0.301\\
14 & 0.899 & 0.726 & 0.780 & 0.772 & 0.397\\
16 & 0.847 & 0.698 & 0.769 & 0.720 & 0.318\\
\end{tabular}
\end{ruledtabular}
\end{table}

\section{Frozen-budget replay}
\label{app:budget-replay}

\begin{table}[H]
\caption{\label{tab:budget-replay} Frozen-score budget replay at the primary depth ($\delta=0$). Reference-defined upper-quartile allocation and uniform allocation receive equal numbers of continuation evaluations. The equal-size row averages event probabilities over $n=8$--$16$ before computing the yield ratio and evaluations per hit. All intervals use the paired policy bootstrap, consistently with Table~\ref{tab:prefix-baselines}. Additional fixed baselines and exact-rank sensitivity analyses are reported in Appendix~\ref{app:prefix-baseline-sensitivity}.}
\begin{ruledtabular}
\begin{tabular}{crrrr}
Scope & Top-quartile event probability & Uniform event probability & Yield ratio [95\% CI] & Evaluations per hit, top/uniform\\
\hline
$n=8$ & 0.0020522 & 0.0009864 & 2.080 [2.016--2.148] & 487.3/1013.7\\
$n=10$ & 0.0018461 & 0.0009959 & 1.854 [1.797--1.913] & 541.7/1004.1\\
$n=12$ & 0.0018559 & 0.0010100 & 1.838 [1.789--1.888] & 538.8/990.1\\
$n=14$ & 0.0016412 & 0.0009905 & 1.657 [1.614--1.702] & 609.3/1009.6\\
$n=16$ & 0.0015457 & 0.0009952 & 1.553 [1.503--1.604] & 646.9/1004.8\\
\hline
Equal-$n$ & 0.0017882 & 0.0009956 & 1.796 [1.772--1.821] & 559.2/1004.4\\
\end{tabular}
\end{ruledtabular}
\end{table}

\section{Fixed-policy baseline comparison and sensitivity}
\label{app:prefix-baseline-sensitivity}

At the primary depth, the seven-component score selects prefixes with higher fresh-event probability than the fixed probability/history (PH) comparator at every size (Table~\ref{tab:prefix-baseline-size}). The relative gain is about $30\%$ at $n=8$, $10\%$ at $n=14$, and $6.8\%$ at $n=16$. The equal-size direct difference is the primary contrast of this baseline comparison. Policy definitions and paired resampling are specified in Appendix~\ref{app:baseline-policy-comparison}.

\begin{table}[H]
\caption{\label{tab:prefix-baseline-size} Full-score and probability/history (PH) allocation at the reference-defined upper-quartile cutoff. Both ranked policies select $494$, $508$, $533$, $528$, and $246$ prefixes at $n=8,10,12,14,16$, respectively. Event rates and their difference are reported per $10\,000$ continuation evaluations; brackets give pointwise $95\%$ intervals. The equal-size row averages probabilities over all five sizes before forming contrasts.}
\centering
\small
\setlength{\tabcolsep}{5pt}
\begin{ruledtabular}
\begin{tabular}{lrrrr}
Scope & Full rate & PH rate & Full $-$ PH [95\% CI] & Full/PH [95\% CI]\\
\hline
$n=8$ & 20.522 & 15.822 & 4.700 [3.981--5.471] & 1.297 [1.244--1.358]\\
$n=10$ & 18.461 & 14.923 & 3.538 [2.972--4.136] & 1.237 [1.194--1.285]\\
$n=12$ & 18.559 & 15.508 & 3.051 [2.516--3.603] & 1.197 [1.159--1.238]\\
$n=14$ & 16.412 & 14.875 & 1.537 [1.202--1.883] & 1.103 [1.079--1.129]\\
$n=16$ & 15.457 & 14.478 & 0.979 [0.584--1.356] & 1.068 [1.039--1.096]\\
\hline
Equal-$n$ & 17.882 & 15.121 & 2.761 [2.527--3.010] & 1.183 [1.165--1.201]\\
\end{tabular}
\end{ruledtabular}
\end{table}

To examine sensitivity to the selection fraction, we rank each score-based policy and, at $n=8$--$14$, select exactly $205$, $512$, or $1024$ prefixes at the original $\gamma_n$. The exact-$512$ rule differs from the original reference-defined upper-quartile policy. At $n=16$, the corresponding exact counts are $100$, $250$, and $500$. For threshold sensitivity, the exact-$512$ selections ($250$ at $n=16$) are held fixed while the event threshold is multiplied by $0.9$ or $1.1$; the $1.0\gamma_n$ case is the top-$25\%$ row. These are changes in the absolute threshold, without recalibrating its quantile.

\begin{figure}[H]
\centering
\includegraphics[width=0.98\textwidth]{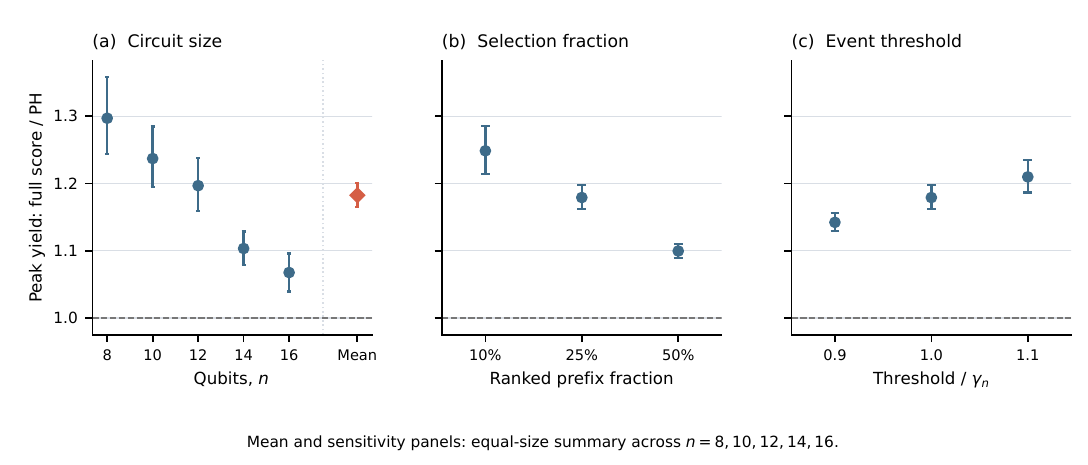}
\caption{\label{fig:prefix-baseline-sensitivity} \textbf{The full-score advantage persists across the tested selection and threshold choices.} Ratios compare the seven-component score with the probability/history (PH) rule. (a) Reference-defined upper-quartile selection by size; the diamond gives the five-size equal-weight result. (b) Exact ranked selection of approximately $10\%$, $25\%$, or $50\%$ of prefixes. (c) Absolute threshold factors with top-$25\%$ selections held fixed. In (b,c), probabilities are averaged equally across all five sizes before forming ratios. Bars are pointwise $95\%$ paired intervals; dashed lines mark equal yield.}
\end{figure}

The full-score advantage persists throughout these checks (Table~\ref{tab:prefix-baseline-sensitivity}; Fig.~\ref{fig:prefix-baseline-sensitivity}). Every size-specific full-minus-PH interval remains above zero in the six evaluated conditions, including the primary policy. The checks reuse the same prefixes and A/B banks; their pointwise intervals describe robustness within these data rather than independent replications.

\begin{table}[H]
\caption{\label{tab:prefix-baseline-sensitivity} Sensitivity of the full-score advantage over PH, with equal weight to $n=8$--$16$. Selection-fraction rows use the original $\gamma_n$; threshold rows hold each policy's top-$25\%$ selection fixed. Rates and differences are per $10\,000$ continuation evaluations. Brackets give pointwise $95\%$ paired intervals. All policies, size-specific results, selection counts, and bank event counts are included in the accompanying analysis records.}
\centering
\small
\setlength{\tabcolsep}{5pt}
\begin{ruledtabular}
\begin{tabular}{lrrrr}
Condition & Full rate & PH rate & Full $-$ PH [95\% CI] & Full/PH [95\% CI]\\
\hline
Top $10\%$ & 23.487 & 18.811 & 4.676 [4.117--5.257] & 1.249 [1.215--1.285]\\
Top $25\%$ & 17.871 & 15.156 & 2.715 [2.477--2.966] & 1.179 [1.162--1.198]\\
Top $50\%$ & 14.001 & 12.733 & 1.268 [1.149--1.388] & 1.100 [1.090--1.110]\\
$0.9\gamma_n$ & 58.514 & 51.229 & 7.285 [6.678--7.913] & 1.142 [1.129--1.156]\\
$1.1\gamma_n$ & 5.869 & 4.851 & 1.018 [0.920--1.123] & 1.210 [1.187--1.235]\\
\end{tabular}
\end{ruledtabular}
\end{table}

\section{Dataset and continuation census}
\label{app:census}

\begin{table}[H]
\caption{\label{tab:census} Dataset and continuation census. A/B are shared-column banks and C/D are row-specific banks. Additional depth-specific continuation and phase allocations are given in Appendix~\ref{app:native-depthgrid}.}
\begin{ruledtabular}
\begin{tabular}{lccccc}
Study component & $n$ & RN & RP & States/$n$ & Continuation unit\\
\hline
Structure and exact interference & 8,10,12,14,16 & 1000 & 1000 & --- & ---\\
Depth-robustness structure & 8,10,12,14 & $1000$/depth & $1000$/depth & --- & ---\\
Susceptibility A/B & 8,10,12,14 & --- & --- & 2048 & $16\,384$ shared columns/bank\\
Susceptibility C/D & 8,10,12,14 & --- & --- & 2048 & $16\,384$ replicates/state/bank\\
Selected-state replacement A/B & 8,10,12,14 & 1000 & 1000 & 2000 & $16\,384$ shared columns/bank\\
Double-held-out phase intervention A/B & 10,14 & 744 & 744 & 1488 & $2048$ held-out columns/bank\\
Cross-size phase extension A/B & 8,12 & 744 & 744 & 1488 & $2048$ fresh columns/bank\\
Cross-size phase extension A/B & 16 & 1000 & 1000 & 2000 & First $2048$ columns/bank\\
Selected-state replacement A/B & 16 & 1000 & 1000 & 2000 & $16\,384$ shared columns/bank\\
Score prediction and replay A/B & 16 & --- & --- & 1000 & Same RN states and A/B banks\\
\end{tabular}
\end{ruledtabular}
\end{table}

Construction and feature-ablation studies use paired trials, whereas gate-deformation tests use anchor states. Sample allocations and resampling schemes are detailed in Appendices~\ref{app:construction} and \ref{app:deformation}, respectively.

State subsets were selected by an outcome-blind hash rule. At $n=10,14$, confirmation used states, suffix columns, and phase seeds disjoint from the initial $256$-state-per-ensemble experiment; the state rule, thresholds, and primary estimand were frozen before the held-out outcomes were evaluated. Each selected state received eight fresh phase replicas. At $n=16$, the phase experiment uses all $1000$ states per ensemble and the first $2048$ fixed columns from each existing A/B continuation bank, shared across ensembles and intervention conditions. Coherent outcomes are reused, and eight fresh phase replicas per state are paired across A/B. The independently calibrated threshold $\gamma_{16}$ is held fixed. Table~\ref{tab:census} gives the state and continuation counts, including the cross-size extensions at $n=8,12,16$; exact index partitions are recorded in the accompanying \nolinkurl{phase_partition_notes.md}.

\raggedbottom
\section{Trajectory-guided construction and feature ablation}
\label{app:construction}

\subsection{Reference information and layerwise selection}

At $\delta=0$, construction reuses existing frozen primary-depth RP reference sets. For $\delta=1,2,3$, each library contains $256$ accepted RP circuits selected by hash ranks of circuit IDs from the matching depth-production pools that also supply the robustness study. Reference IDs and observable trajectories are fixed before the new construction trials.

The construction objective uses
\begin{equation}
\boldsymbol f_\ell=
\left(\ln\GammaPeak(|\Psi_\ell\rangle),\ \widetilde H_1(p_\ell),\
\ln\!\left[2^n\sum_zp_\ell(z)^2\right],\ R_\ell,\
\overline I_Q^{\mathrm{adj}},\ \overline E_{\mathrm{cut}},\
\overline\lambda_{\max}\right),
\label{eq:construction-features}
\end{equation}
where $\overline I_Q^{\mathrm{adj}}$ averages QMI over neighboring pairs and $\overline\lambda_{\max}$ averages the largest squared Schmidt coefficient over contiguous cuts. Unlike the pre-final redistribution summary, the construction evaluates $R_1$ relative to $p_0(z)=\mathbf1\{z=0^n\}$. All features are recomputed on each candidate trajectory. The normalization is $s_{\ell k}=\max\{\operatorname{SD}_{\mathrm{RN}}(f_{\ell k}),0.05\}$ from the matched reference library. This vector includes instantaneous peak information and is not the seven-component prefix score or the elastic-net classifier.

Each trial draws an RP reference uniformly from its fixed library and evaluates $32$ Haar candidate blocks per layer. The four candidates with smallest current loss form $\mathcal S_\ell$. One-layer lookahead selects the smallest
\begin{equation}
\mathcal J_\ell(j)=\frac12\left[\mathcal L_\ell(j)+
\min_{b=1,\ldots,4}\mathcal L_{\ell+1}(j,b)\right],
\qquad j\in\mathcal S_\ell.
\label{eq:construction-lookahead}
\end{equation}
The next-layer blocks are shared across the four shortlisted candidates. Only the current selected block is committed, and the last layer uses the immediate loss. Targets are not adaptively reselected. Proposal banks are shared across depths within each size and trial, and each size--depth cell uses its matched RP library.

\subsection{Common adaptive refinement and complete generation results}

At each of four rounds, we evaluate candidate updates to the current circuit. The candidates combine three supports (one gate, a connected pair, or four gates), $16$ normalized random directions per support family, four step sizes $0.01,0.03,0.1,0.3$, and both signs, giving $3\times16\times4\times2=384$ candidates per round. For a gate $G_g$, the update is
\begin{equation}
G_g'=\exp\!\left(-i\sum_{a=1}^{15}\theta_{ga}P_a\right)G_g,
\label{eq:construction-gate-update}
\end{equation}
with $P_a$ the nonidentity two-qubit Pauli products divided by two. The incumbent and new candidates are ranked first by $\min\{\GammaPeak/\gamma_n^{\mathrm{search}},1\}$, then by smallest phase-aligned terminal-state distance to the initial circuit, $\sqrt{2-2|\langle\psi_{\mathrm{initial}}|\psi\rangle|}$. The fixed refinement targets are $\gamma_n^{\mathrm{search}}=30.0,31.5,33.0,35.1$ for $n=8,10,12,14$, respectively (Table~\ref{tab:threshold-roles}). At $n=12$, the reported success criterion $\gamma_{12}^{\mathrm{RP}}=33.2$ is a saved secondary evaluation of the same generated circuits; it did not change the search target, ranking rule, or generated outcomes. The accompanying \texttt{construction\_protocol.json} records this reporting distinction. Ties use candidate order, with the incumbent first. The selected circuit becomes the next round's base. Thus the procedure is adaptive, but gives no extra peak reward after the threshold has been reached. T+R and H+R use the same rule and proposal directions.

\begin{table}[H]
\caption{\label{tab:construction} Complete construction results. Counts are successes at the RP thresholds $\gamma_n^{\mathrm{RP}}$ in Eq.~\eqref{eq:rp-eligibility}, out of $256$ trials per arm and setting. T and H denote trajectory and Haar initialization; +R adds the common $1536$-candidate refinement. The final column is T+R minus H+R in percentage points with paired $95\%$ intervals.}
\begin{ruledtabular}
\begin{tabular}{ccrrrrr}
$n$ & $\delta$ & T & H & T+R & H+R & Gain [95\% CI]\\
\hline
8 & 0 & 0 & 0 & 22 & 1 & 8.20 [4.69, 12.11]\\
8 & 1 & 0 & 0 & 16 & 0 & 6.25 [3.52, 9.38]\\
8 & 2 & 0 & 0 & 50 & 0 & 19.53 [14.84, 24.61]\\
8 & 3 & 0 & 0 & 45 & 0 & 17.58 [12.89, 22.27]\\
10 & 0 & 0 & 0 & 9 & 0 & 3.52 [1.56, 5.86]\\
10 & 1 & 0 & 0 & 36 & 0 & 14.06 [9.77, 18.36]\\
10 & 2 & 0 & 0 & 48 & 0 & 18.75 [14.05, 23.83]\\
10 & 3 & 2 & 0 & 117 & 2 & 44.92 [39.06, 50.79]\\
12 & 0 & 0 & 0 & 9 & 0 & 3.52 [1.56, 5.86]\\
12 & 1 & 0 & 0 & 9 & 0 & 3.52 [1.56, 5.86]\\
12 & 2 & 0 & 0 & 47 & 1 & 17.97 [13.28, 22.67]\\
12 & 3 & 0 & 0 & 39 & 1 & 14.84 [10.55, 19.53]\\
14 & 0 & 0 & 0 & 1 & 0 & 0.39 [0.00, 1.17]\\
14 & 1 & 0 & 0 & 18 & 0 & 7.03 [3.91, 10.16]\\
14 & 2 & 0 & 0 & 19 & 0 & 7.42 [4.30, 10.55]\\
14 & 3 & 0 & 0 & 75 & 0 & 29.30 [23.83, 35.16]\\

\end{tabular}
\end{ruledtabular}
\end{table}

Target-cluster resampling retains all trials assigned to each sampled RP target; its $95\%$ intervals remain positive in the $12$ settings at $n=8,10,12$. The paired-trial intervals in Table~\ref{tab:construction} are nominal and setting-specific. Generation, target selection, and refinement were frozen before trial outcomes were evaluated.

\subsection{Intermediate-feature ablation across size and depth}
\label{app:ablation-design}

At each size--depth setting, the multivariate-versus-middle-peak-only ablation uses $256$ shared proposal seeds and two fixed, matched reference packages, each containing $256$ RP and $256$ RN trajectories; package membership is disjoint within each ensemble and setting. Reference populations are reused, while construction proposals and fresh suffixes are independently generated for each setting. Each seed and package selects a common RP target for the two arms. Both arms compute all seven diagnostics on the same $32$ candidate blocks per layer, but middle-peak-only uses only $\ln\GammaPeak$ in the loss at layers $2$ through $L_\delta-1$. Targets and feature masks remain identical at the first and last layers. Greedy selection uses neither lookahead nor terminal refinement.

All generated trials are retained. Their intermediate states at $t=n$ are evaluated with $2048$ new common suffixes following the setting's reverse-suffix schedule. Terminal and fresh events use the fixed $\gamma_n$ in Table~\ref{tab:threshold-roles}. Reference-specific arm contrasts are averaged within seed before resampling. Terminal intervals use $10\,000$ paired-seed bootstrap replicates; fresh intervals use $2000$ paired seed--suffix replicates, preserving common columns. Both reference packages have positive fresh contrasts in every setting. Table~\ref{tab:crosssize-ablation} reports all $16$ settings separately; \nolinkurl{depthgrid_ablation_protocol.json} provides the protocol and source-result hashes.

\begin{table}[H]
\caption{\label{tab:crosssize-ablation} Intermediate-feature ablation across size and depth, $L_\delta=3n/2-\delta$. M denotes multivariate guidance and P denotes middle-peak-only guidance. Terminal probabilities are percentages; fresh probabilities at cut $t=n$ are reported per $10\,000$ continuations. Differences are M minus P in the corresponding units, with nominal $95\%$ intervals. Each arm uses $256$ trials per reference package and setting; inference averages the two reference contrasts within each paired seed.}
\begingroup
\small
\setlength{\tabcolsep}{4pt}
\begin{ruledtabular}
\begin{tabular}{ccrrrrrr}
& & \multicolumn{3}{c}{Terminal (\%)} & \multicolumn{3}{c}{Fresh (per $10\,000$)}\\
$n$ & $\delta$ & M & P & Difference [95\% CI] & M & P & Difference [95\% CI]\\
\hline
8 & 0 & 10.55 & 3.71 & 6.84 [3.71, 10.16] & 21.458 & 11.997 & 9.46 [6.89, 11.82]\\
 & 1 & 8.01 & 2.54 & 5.47 [2.73, 8.40] & 24.862 & 12.407 & 12.45 [9.79, 15.23]\\
 & 2 & 21.09 & 11.72 & 9.38 [5.08, 13.67] & 95.549 & 49.667 & 45.88 [37.07, 54.61]\\
 & 3 & 26.37 & 10.74 & 15.63 [10.94, 20.31] & 119.696 & 44.069 & 75.63 [65.29, 86.16]\\[3pt]
10 & 0 & 7.81 & 2.54 & 5.27 [2.73, 7.81] & 18.063 & 11.673 & 6.39 [4.31, 8.57]\\
 & 1 & 16.80 & 9.18 & 7.62 [3.32, 11.91] & 63.314 & 38.404 & 24.91 [19.12, 30.41]\\
 & 2 & 16.60 & 9.96 & 6.64 [2.93, 10.55] & 61.750 & 38.118 & 23.63 [18.84, 28.80]\\
 & 3 & 41.21 & 29.49 & 11.72 [5.66, 17.77] & 231.180 & 146.332 & 84.85 [68.77, 101.17]\\[3pt]
12 & 0 & 5.27 & 2.34 & 2.93 [0.59, 5.47] & 16.384 & 11.072 & 5.31 [3.35, 7.31]\\
 & 1 & 4.69 & 2.54 & 2.15 [0.00, 4.30] & 16.956 & 10.967 & 5.99 [3.99, 8.05]\\
 & 2 & 17.97 & 10.55 & 7.42 [3.13, 11.72] & 61.436 & 44.041 & 17.40 [12.37, 22.35]\\
 & 3 & 18.95 & 14.26 & 4.69 [0.39, 8.98] & 69.456 & 44.794 & 24.66 [18.89, 30.51]\\[3pt]
14 & 0 & 3.91 & 4.69 & -0.78 [-3.13, 1.56] & 13.723 & 10.338 & 3.39 [1.61, 5.09]\\
 & 1 & 15.04 & 10.94 & 4.10 [0.00, 8.01] & 43.688 & 31.843 & 11.84 [8.18, 15.36]\\
 & 2 & 14.45 & 8.79 & 5.66 [1.95, 9.57] & 46.644 & 35.419 & 11.22 [7.07, 15.22]\\
 & 3 & 40.63 & 30.47 & 10.16 [4.30, 16.02] & 181.170 & 120.068 & 61.10 [49.39, 73.26]\\

\end{tabular}
\end{ruledtabular}
\endgroup
\end{table}

\subsection{Original single-size feature and ordering controls}

The original $n=8$, $L=12$ feature ablation uses the greedy selection and sample allocations of Appendix~\ref{app:ablation-design}, with separate proposal seeds and fresh suffixes. Ordered, middle-shuffled, middle-peak-only, early-multifeature, and late-multifeature policies are evaluated with both the full feature vector and the vector excluding redistribution. First- and last-layer targets and feature masks remain fixed. Middle-peak-only uses only $\ln\GammaPeak$ at layers $2$--$11$; early and late policies restore the other features at layers $2$--$6$ or $7$--$11$, respectively. The shuffled policy permutes intermediate targets, not their RN scales. Fixing endpoint targets does not fix the realized endpoint states.

Terminal and fresh events use the calibrated $\gamma_8$ in Table~\ref{tab:threshold-roles}. Fresh suffixes are generated after the construction policies are fixed and applied at cut $t=8$. Pooled contrasts use the same within-seed averaging and paired bootstrap scheme as Appendix~\ref{app:ablation-design}. All planned feature-family contrasts are reported in Table~\ref{tab:construction-ablation-contrasts}.

The ordered-minus-peak-only, ordered-minus-shuffled, and early-minus-late contrasts test feature content, target ordering, and the placement of multivariate guidance, respectively.

Terminal feature matching chooses among the $32$ complete candidate circuits using only their terminal observable distance. Terminal-only search instead chooses layer blocks using simulated terminal peaks; its candidate-gate application count matches the greedy trackers, but its information access differs. Random best of $32$ selects the largest terminal peak among the $32$ Haar circuits. These controls separate intermediate multivariate information, temporal ordering, and direct terminal search.

\begin{table}[H]
\caption{\label{tab:construction-ablation} Feature-ablation and baseline outcomes at $n=8$, grouped into full-feature trackers, trackers without redistribution, terminal feature matching, and common search baselines. Terminal counts are out of $256$ trials for each reference package. The last column is the fresh-event probability at cut $t=8$, averaged over the two packages and reported per $10\,000$ continuations. The three common baselines are evaluated once per shared seed, not once per reference package.}
\begin{ruledtabular}
\begin{tabular}{lrrr}
Feature family / policy & Terminal hits: ref. 1 & Terminal hits: ref. 2 & $10^4q$\\
\hline
Full: ordered & 29 & 31 & 21.887\\
Full: middle peak only & 9 & 6 & 11.511\\
Full: middle shuffled & 28 & 20 & 13.638\\
Full: early multifeature & 11 & 14 & 19.388\\
Full: late multifeature & 20 & 27 & 15.087\\
\hline
No redistribution: ordered & 42 & 27 & 25.043\\
No redistribution: middle peak only & 9 & 5 & 11.625\\
No redistribution: middle shuffled & 29 & 30 & 13.008\\
No redistribution: early multifeature & 15 & 20 & 20.895\\
No redistribution: late multifeature & 29 & 22 & 15.192\\
\hline
Terminal feature match & 6 & 6 & 20.647\\
\hline
Terminal-only search & 13 & --- & 11.387\\
Haar & 0 & --- & 9.136\\
Random best of 32 & 8 & --- & 14.515\\

\end{tabular}
\end{ruledtabular}
\end{table}

\begin{table}[H]
\caption{\label{tab:construction-ablation-contrasts} Planned intermediate-feature contrasts. Terminal differences are percentage points; fresh-event differences are events per $10\,000$. Brackets denote nominal $95\%$ intervals.}
\begin{ruledtabular}
\begin{tabular}{llrr}
Feature family & Contrast & Terminal difference [95\% CI] & Fresh difference [95\% CI]\\
\hline
Full & Ordered minus middle peak only & 8.79 [5.86, 11.91] & 10.38 [8.12, 12.65]\\
Full & Ordered minus middle shuffled & 2.34 [-1.37, 6.05] & 8.25 [6.03, 10.59]\\
Full & Early minus late & -4.30 [-7.42, -1.37] & 4.30 [1.89, 6.67]\\
No redistribution & Ordered minus middle peak only & 10.74 [7.62, 14.06] & 13.42 [10.89, 16.09]\\
No redistribution & Ordered minus middle shuffled & 1.95 [-1.76, 5.86] & 12.04 [9.54, 14.62]\\
No redistribution & Early minus late & -3.12 [-6.25, 0.00] & 5.70 [3.19, 8.23]\\

\end{tabular}
\end{ruledtabular}
\end{table}

The intermediate multivariate contribution is positive in both feature families. Temporal shuffling attenuates fresh susceptibility, but the corresponding terminal intervals include zero.

\flushbottom
\section{Functional stability under finite gate deformations}
\label{app:deformation}

At $n=10,12,14$, a separate outcome-blind selection supplies $128$ RP and $128$ RN anchors per size. All gates in the last completed prefix layer are perturbed along eight random $\mathrm{SU}(4)$ directions per anchor. Their common radius is
\begin{equation}
d_{\mathrm{gate}}
=\left[\sum_{g\in\mathrm{last\ prefix\ layer}}
\left(1-\frac{|\operatorname{Tr}(G_g^\dagger G_g')|^2}{16}\right)\right]^{1/2}
=\sqrt{0.03}.
\label{eq:deformation-radius}
\end{equation}
Directions are sampled before continuation evaluation and averaged, not selected for high outcomes. Each original or deformed prefix receives $16\,384$ common fresh suffixes from an independent bank. Computational-basis phase scrambling is sampled separately for each anchor--direction--suffix evaluation, preserving populations. The event thresholds are the calibrated values in Table~\ref{tab:threshold-roles}. This follow-up uses its own anchors and banks; its absolute rates are not the estimates in Fig.~\ref{fig:phase-intervention}.

\begin{figure}[H]
\centering
\includegraphics[width=0.88\textwidth]{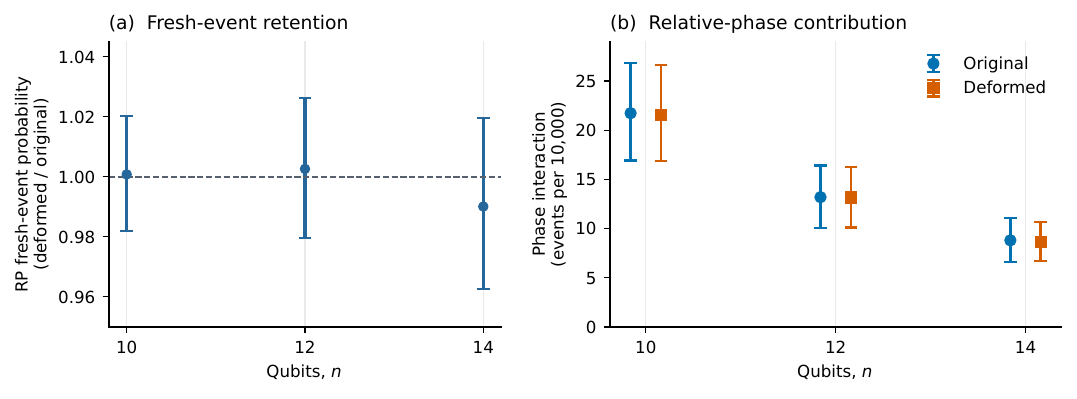}
\caption{\label{fig:deformation} \textbf{RP functionality persists under finite prefix-gate changes.} (a) RP fresh-event probability after deformation divided by its original value; the dashed line denotes unity. At $n=12$, coherent retention follows from boundary Haar invariance. (b) Original and deformed RN--RP phase interactions, in events per $10\,000$ fresh continuations. Filled blue circles denote original states and filled vermilion squares denote deformed states. Points average over the $128$ anchors per ensemble and eight deformation directions. Bars show $95\%$ anchor--suffix bootstrap intervals. The original/deformed markers are offset horizontally for readability.}
\end{figure}

Figure~\ref{fig:deformation}a shows deformed/original RP event-probability ratios of $1.001$ [$0.982,1.020$], $1.003$ [$0.980,1.026$], and $0.990$ [$0.962,1.020$] at $n=10,12,14$. At $n=12$, the deformed last prefix layer and the first fresh-suffix layer share an edge matching, so coherent-probability retention follows from Haar invariance. This boundary symmetry does not apply to the tested $n=10,14$ deformations and does not by itself preserve the phase-scrambled interaction. Figure~\ref{fig:deformation}b applies the phase interaction in Eq.~\eqref{eq:phase-interaction} before and after deformation, averaging over anchors and directions. The deformed interactions remain positive at each size. Inference resamples RN and RP anchors independently and suffix columns jointly across arms, using the existing $5000$-replicate confirmation intervals.

\section{Structural scores and statistical estimators}
\label{app:score-statistics}

\subsection{Reference standardization and repeatability}
\label{app:repeatability}

Let $x_{nij}$ denote component $j$ for prefix $i$ at size $n$, in the order given in Sec.~\ref{subsec:continuations}, and fix the RP-associated orientation vector as $\boldsymbol d=(+1,+1,+1,-1,+1,-1,+1)$. With the mean and population standard deviation from an independent $10\,000$-state reference bank,
\begin{align}
u_{nij}
&=d_j\frac{x_{nij}-\mu_{n,j}^{\mathrm{ref}}}
{\sigma_{n,j}^{\mathrm{ref}}},\\
r_{ni}&=\frac{1}{7}\sum_{j=1}^{7}u_{nij}.
\label{eq:prefix-component-score}
\end{align}
A final reference standardization gives
\begin{equation}
s_{\mathrm{pre},ni}
=\frac{r_{ni}-\mu_{r,n}^{\mathrm{ref}}}
{\sigma_{r,n}^{\mathrm{ref}}}.
\label{eq:prefix-score}
\end{equation}
The feature set and RP-associated directions are fixed before fresh-continuation evaluation. Score evaluation then uses only prefix observables, without using the terminal or regenerated outcomes of the evaluated circuit. The frozen metric order, directions, and reference moments are recorded in \nolinkurl{prefix_score_reference.json} and \nolinkurl{n16_fresh_continuation_reference.json}. The separate classifier record is described in Appendix~\ref{app:classifier-specification}.

For the upper and lower reference-score quartiles, let $\widehat q_{g,n}^{(b)}$ be the event rate in group $g\in\{\mathrm{upper},\mathrm{lower}\}$ at size $n$ and bank $b$. We average event rates with equal weight over $n\in\{8,10,12,14,16\}$ before forming the upper-to-lower risk ratio. Bank-specific summaries use $\overline q_g^{(b)}=\tfrac15\sum_n\widehat q_{g,n}^{(b)}$; the two-bank summary averages $\overline q_g^{(A)}$ and $\overline q_g^{(B)}$ before taking their group ratio. Reference cutoffs and score definitions remain fixed. Intervals use $10\,000$ multiway bootstrap draws: prefixes are resampled within each quartile group with common weights across $A/B$, and suffix columns are resampled independently within each bank with common weights across groups. Event rates use the product of prefix and suffix weights. Independent size-specific draws are averaged with the same weights, and percentile bounds are then computed for each ratio. The primary RP/RN comparison uses the same rate-first, equal-size aggregation.

At $n=8,10,12,14$, repeatability across independently drawn continuations is measured with the two row-specific banks $C$ and $D$. For $N=2048$ prefixes and $R=16\,384$ continuations per bank, define
\begin{align}
\widehat q_{ni}^{(b)}
&=\frac{1}{R}\sum_{r=1}^{R}
\mathbf 1\!\left\{\Gamma_{\mathrm{peak},nir}^{(b)}\geq\gamma_n\right\},
\qquad b\in\{C,D\},\\
V_n
&=\frac{1}{N-1}\sum_{i=1}^{N}
(\widehat q_{ni}^{(C)}-\overline q_n^{(C)})
(\widehat q_{ni}^{(D)}-\overline q_n^{(D)}),\\
V_G&=\frac14\sum_{n\in\{8,10,12,14\}}V_n.
\label{eq:repeatable-propensity}
\end{align}
Here $\overline q_n^{(b)}=N^{-1}\sum_i\widehat q_{ni}^{(b)}$. Because the two banks are independent, continuation-sampling noise has zero cross-covariance; $V_n$ therefore estimates repeatable between-prefix propensity variance, in probability-squared units.

The equal-size estimate is $V_G=5.669\times10^{-7}$ probability squared [$95\%$ CI: $5.111\times10^{-7}$--$6.292\times10^{-7}$].

\subsection{Fixed-policy baseline comparison}
\label{app:baseline-policy-comparison}

At the primary depth, we compare six allocation rules on unconditional prefixes at all five sizes: $2048$ prefixes per size for $n=8$--$14$ and $1000$ at $n=16$, with shared banks $A$ and $B$ each containing $M=16\,384$ continuations. The rules are uniform allocation, high prefix $\GammaPeak$, low normalized Shannon entropy, high collision ratio, a probability/history (PH) score, and the original seven-component score. For $X\in\{R,H,C\}$, let $z_{X,ni}=(X_{ni}-\mu_{X,n}^{\mathrm{ref}})/\sigma_{X,n}^{\mathrm{ref}}$, using the original independent reference moments. Here $R$ is redistribution at the last prefix transition, $H=\widetilde H_1$ at the cut, and $C$ is the collision ratio. Define
\begin{equation}
s_{\mathrm{PH},ni}=\frac{z_{R,ni}-z_{H,ni}+z_{C,ni}}{3}.
\label{eq:probability-history-score}
\end{equation}
The comparator definitions, directions, selection rules, and primary contrast were fixed before computing the comparative replay results. No weights or reference scales were fitted to those results.

The primary full-score set retains $s_{\mathrm{pre},ni}$ at or above the original reference $0.75$ quantile, selecting $494$, $508$, $533$, $528$, and $246$ prefixes at $n=8,10,12,14,16$. Each ranked comparator selects the same count, preferring lower entropy and higher values of the other rankings; ties use ascending fixed prefix ID. Uniform allocation uses all prefixes. For policy $a$ with selected set $S_{a,n}$, the event rate and equal-size summary are
\begin{align}
\widehat q_{a,n}^{(b)}
&=\frac{1}{M|S_{a,n}|}\sum_{i\in S_{a,n}}\sum_{r=1}^{M}
Y_{nibr},\\
\widehat q_a
&=\frac15\sum_{n\in\{8,10,12,14,16\}}
\frac{\widehat q_{a,n}^{(A)}+\widehat q_{a,n}^{(B)}}{2},
\label{eq:baseline-policy-risk}
\end{align}
where $Y_{nibr}=\mathbf1\{\Gamma_{\mathrm{peak},nibr}\geq\gamma_n\}$. Differences and ratios are formed after averaging probabilities. The primary contrast is $10^4(\widehat q_{\mathrm{full}}-\widehat q_{\mathrm{PH}})$; the full-to-PH ratio accompanies it, and all other policy comparisons are secondary.

Intervals use $10\,000$ paired additive prefix--suffix bootstrap replicates (seed $20261025$ for $n=8$--$14$ and $20261026$ for $n=16$). Within each size, prefixes are resampled within the original full-score selected and complementary strata, using common draws across banks, policies, and sensitivity conditions. Each policy's prefix-resampled rate uses its own number of selected resampled rows. Suffix columns are resampled independently within $A$ and $B$, with common column draws across policies and conditions. The centered mean of the two bank-specific suffix-resampled rates is added to the prefix-resampled rate, preserving policy overlap and shared-bank dependence. Size-specific resamples are independent; their probabilities are averaged before computing each contrast. Selection sets and reference quantities remain fixed. The $95\%$ percentile intervals are pointwise; only the main equal-size full-minus-PH contrast is the primary endpoint. This is a conditional retrospective comparison on the existing prefix and continuation banks.

For selection-fraction sensitivity, all ranked policies select approximately $10\%$, exactly $25\%$, or exactly $50\%$ of the cohort at the original $\gamma_n$: $205$, $512$, or $1024$ prefixes for $n=8$--$14$, and $100$, $250$, or $500$ at $n=16$. The exact-$25\%$ rule is distinct from the primary reference-quartile rule. Threshold sensitivity fixes this exact-$25\%$ count and uses $0.9\gamma_n$ and $1.1\gamma_n$, without quantile recalibration. The $1.0\gamma_n$ condition is the exact-$25\%$ selection result. All conditions reuse the same prefixes and banks. Results appear in Appendix~\ref{app:prefix-baseline-sensitivity}; the complete rules, masks, numerical summaries, and bootstrap draws accompany the analysis records.

\subsection{Effect sizes and resampling}

Independent RN--RP effects use
\begin{align}
g_M
&=J\,
\frac{\overline S_{\mathrm{RP},M}^{\mathrm{pre}}
-\overline S_{\mathrm{RN},M}^{\mathrm{pre}}}{s_{p,M}},\\
J&=1-\frac{3}{4(N_{\mathrm{RP}}+N_{\mathrm{RN}})-9},
\label{eq:hedges}
\end{align}
where $s_{p,M}$ is the pooled sample standard deviation \cite{Hedges1981}. Positive $g$ denotes a larger RP mean.

Structural and exact-interference intervals use ensemble-stratified circuit bootstraps: $1000$ replicates for primary structural effects at $n=8$--$14$, $5000$ for the $n=16$ and depth extensions, and $10\,000$ for interference. Continuation inference resamples states and suffixes while preserving shared-bank pairing. Primary phase intervals use $10\,000$ additive state--suffix draws with nested phase-replica resampling. At $n=16$, the four two-sided Welch tests of circuit-level pre-final endpoint means form one Holm family \cite{Holm1979}; effect-size intervals remain pointwise. The additive combination, nested resampling, and model-bootstrap weights are specified in \nolinkurl{statistical_inference_notes.md}, alongside the implementations of these nonparametric procedures \cite{EfronTibshirani1993}.

For the RP/RN mean $F_{t_*}^{\mathrm{peak}}$ ratio, RN and RP circuits are resampled independently within each size. Each of $10\,000$ replicates forms the ratio of the bootstrapped RP and RN ensemble means, and percentile intervals are reported.

\subsection{Cross-size classifier specification}
\label{app:classifier-specification}

The $15$ curves comprise classical and quantum pairwise MI at adjacent, short-range, and long-range separations; total correlation per qubit; full-$L_1$ redistribution; cut-averaged and half-cut von Neumann entropy and largest squared Schmidt coefficient; cut-averaged effective rank; normalized output entropy; and collision ratio. After removing the final complete checkpoint, each remaining curve is linearly interpolated on its normalized pre-final interval at ten equally spaced positions from $0$ to $1$ and concatenated in curve-major order.

Standardization is fitted only on the training data. Model selection uses mean AUROC across the two source-size directions, $n=10\to12$ and $12\to10$, before refitting on all $4000$ source circuits. The selected logistic classifier uses the elastic-net family at its $L_1$ endpoint ($C=0.03$, \nolinkurl{l1_ratio=1}), balanced class weights, and the SAGA solver; $13$ coefficients are nonzero. The full representation, training protocol, preprocessing, fitted parameters, and implementation references are provided in \nolinkurl{trajectory_score_150d_reference.json}.

Applied without refitting to the $n=16$ cohort of $1000$ RN and $1000$ RP circuits, the classifier gives an RP-minus-RN trajectory-score Hedges effect of $g=1.220$ [$95\%$ CI: $1.127$--$1.315$].

\begin{samepage}
\subsection{Fresh-event log-odds model}
\label{app:score-logit-model}

For prefix $i$ at size $n$, bank $b$, and shared suffix column $r$, the calibrated event indicator is modeled with conditional probability $\pi_{nibr}$:
\begin{equation}
\operatorname{logit}\pi_{nibr}
=\alpha_n+\kappa_{nb}+\beta_n s_{\mathrm{pre},ni}
+u_{ni}+v_{nbr},
\qquad b\in\{A,B\}.
\label{eq:score-crossed-logit}
\end{equation}
Here $\alpha_n$ is the size-specific intercept and $\kappa_{nA}=0$ fixes the bank baseline. The Gaussian prefix intercept $u_{ni}$ is shared across banks, with a variance shared across $n=8$--$14$; Gaussian suffix intercepts $v_{nbr}$ have size- and bank-specific variances. The cell-level Bernoulli fit uses empirical-Bayes penalized quasi-likelihood, updating variance components by posterior second moments and the common prefix variance with equal weight per size.
\end{samepage}

We report the size-specific slopes $\widehat\beta_n$ per reference-score standard deviation. The $10\,000$-replicate multiway cluster bootstrap resamples prefixes jointly across $A/B$ within each size and suffix columns independently within each size and bank. For $n=8$--$14$, each replicate refits the coefficients and variance components; $95\%$ intervals use the corresponding percentile bounds.

At $n=16$, the prefix-intercept variance is fixed at the $n=8$--$14$ estimate $0.0900702$, while suffix variances are estimated and the coefficients and suffix variances are refitted in all $10\,000$ bootstrap draws (seed $20261026$). Estimating an $n=16$-specific prefix variance gives a sensitivity slope of $0.3760$, compared with $0.3810$ under the fixed variance.

\section{Retention likelihood and fresh-event enrichment}
\label{app:selection}

Let $\nu_{t_*}$ be the intermediate-state law of the random-circuit generator. Let $H_{\mathrm{RP}}^{(0)}$ be the original RP membership event, including its terminal-peak and retention rules, and define $h(\psi)=\Prob(H_{\mathrm{RP}}^{(0)}\mid\Psi_{t_*}=\psi)$. Bayes' rule gives
\begin{equation}
\frac{\dd\nu_{t_*\mid H_{\mathrm{RP}}^{(0)}}}{\dd\nu_{t_*}}(\psi)
=\frac{h(\psi)}{\Prob(H_{\mathrm{RP}}^{(0)})}.
\label{eq:selection-tilt}
\end{equation}
All expectations are taken under $\nu_{t_*}$. For an independently regenerated event $B_n^{(1)}$, fresh-suffix independence ensures $B_n^{(1)}\perp H_{\mathrm{RP}}^{(0)}\mid\Psi_{t_*}$. With $q=q_{n,t_*}(\Psi_{t_*})$ and $h=h(\Psi_{t_*})$,
\begin{equation}
\Prob(B_n^{(1)}\mid H_{\mathrm{RP}}^{(0)})
=\frac{\E[q h]}{\E[h]}.
\label{eq:selected-risk}
\end{equation}
Since $\Prob(B_n^{(1)})=\E[q]$ for the random-circuit ensemble, the enrichment due to RP selection is
\begin{equation}
\Prob(B_n^{(1)}\mid H_{\mathrm{RP}}^{(0)})
-\Prob(B_n^{(1)})
=\frac{\operatorname{Cov}(q,h)}{\E[h]}.
\label{eq:selection-covariance}
\end{equation}
The identity relates enrichment to the association between fresh-event susceptibility and the likelihood of RP selection.

For comparison, suppose selection is solely the same event used for fresh evaluation, the original and regenerated suffixes follow the same law and are conditionally independent given the state, and no rank retention is applied. Then $h=q$, and for $\E[q]>0$,
\begin{equation}
\Prob(B^{(1)}\mid B^{(0)})-\Prob(B^{(1)})
=\frac{\E[q^2]}{\E[q]}-\E[q]
=\frac{\operatorname{Var}(q)}{\E[q]}\geq0.
\label{eq:same-event-enrichment}
\end{equation}
In this special case, between-state susceptibility variation produces enrichment through selection. Our RP and fresh-event criteria differ, so the more general covariance identity applies. The structural, phase, and construction tests establish which state properties accompany and support this enrichment.

\section{Fresh-continuation tests across nominal depths}
\label{app:native-depthgrid}

To assess robustness to depth variation, we extend continuation replacement, population-preserving phase intervention, and prefix-score selection to $\delta=1,2,3$ at $n=8,10,12,14$. Each setting uses its existing $1000$ RN and $1000$ RP states at $t_*=n$. Two independent banks A/B contain $1024$ fresh suffixes each; their columns are shared across both ensembles, the coherent state and its eight phase replicas, and the unconditional score experiment. The latter reuses the original $2048$ prefixes and unchanged seven-component score. Independent row-specific banks C/D each supply $1024$ continuations per prefix. Suffixes follow the remaining-depth law of the corresponding nominal setting. The calibrated primary $\gamma_n$ is held fixed, so baseline event rates may vary with depth. An event is counted only after the complete continuation; a longer suffix does not create additional counted threshold-crossing opportunities. This grid tests persistence across the studied size--depth settings.

Native and score-selection intervals use $10\,000$ additive two-way bootstrap draws: independent prefix weights within each native ensemble or score group, together with shared suffix-column weights within bank. Coherent and scrambled outcomes remain paired, and the eight phase replicas are averaged within state. C/D covariance uses paired-prefix resampling. For the allocation replay, the top $512$ scores are compared with uniform allocation over all $2048$ prefixes. The separate reference-quartile comparison retains the original reference-derived boundaries. Each depth is reported separately; reused prefixes are not counted as independent observations across depths. Intervals are pointwise $95\%$ intervals.

\begin{figure}[H]
\centering
\includegraphics[width=0.98\textwidth]{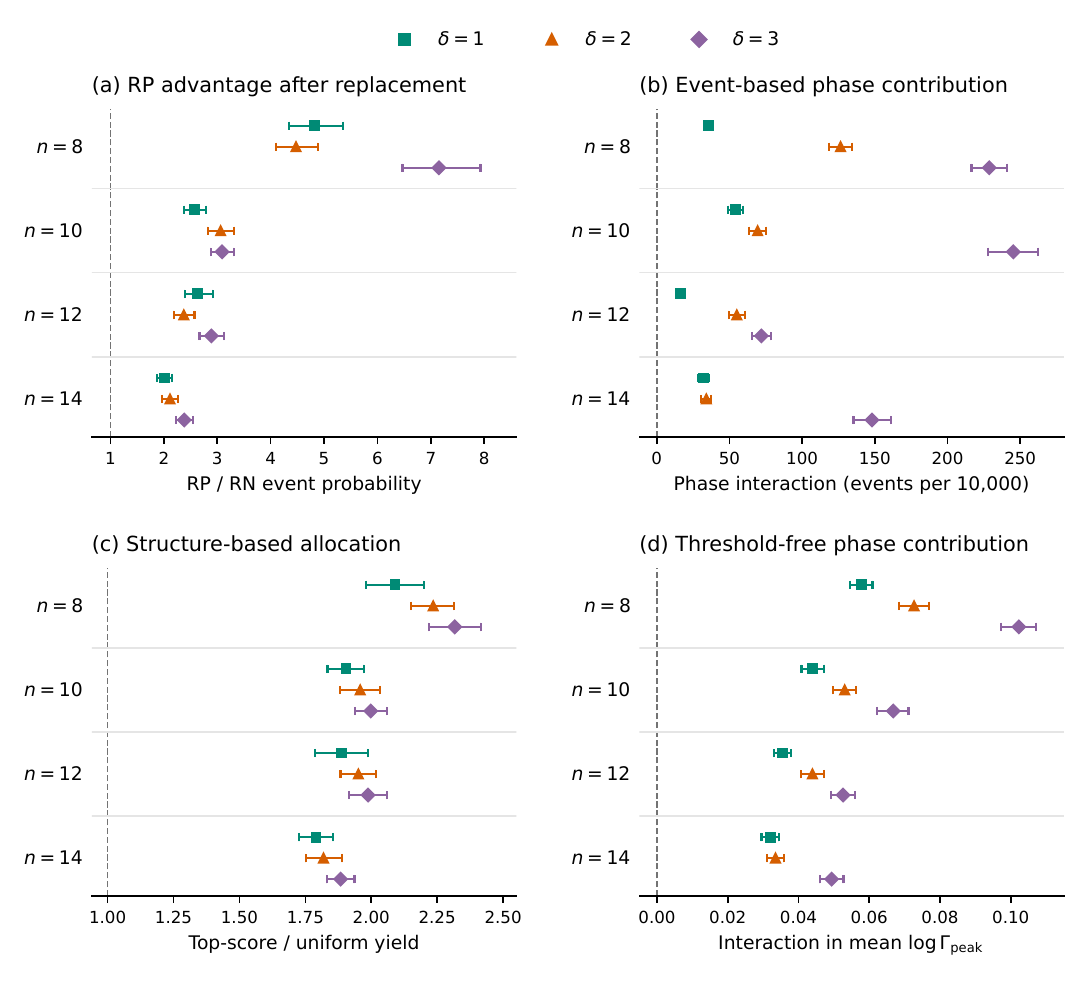}
\caption{\label{fig:native-depthgrid} \textbf{Fresh-continuation effects persist across the added depth settings.} All $12$ combinations of $n=8,10,12,14$ and $\delta=1,2,3$ are shown; within each size, offsets increase from top to bottom. (a) RP/RN fresh-event probability ratios. (b) Event-based phase interactions per $10\,000$ continuations. (c) Top-score/uniform allocation gains using the unchanged prefix score. (d) Threshold-free mean-$\log\GammaPeak$ phase interactions. Bars show the pointwise $95\%$ intervals in Table~\ref{tab:native-depthgrid}; dashed lines mark the null value of one for ratios and zero for interactions. All axes are linear. Depth extensions use fixed primary-calibrated event thresholds and are reported separately from the original $\delta=0$ estimates in Figs.~\ref{fig:selected-law}--\ref{fig:phase-intervention}.}
\end{figure}

Across all $12$ added settings, RP/RN and allocation-ratio intervals exceed one and both phase-interaction intervals exceed zero (Table~\ref{tab:native-depthgrid}; Fig.~\ref{fig:native-depthgrid}). Reference-quartile ratios span $3.78$--$10.44$, with every interval above one. All C/D covariance intervals are positive, confirming repeatable susceptibility without using the score. Each A/B bank separately reproduces the RP advantage, both phase interactions, and the allocation gain.

\begin{table}[H]
\caption{\label{tab:native-depthgrid} Direct fresh-continuation effects in the depth extensions. $R_{\mathrm{RP/RN}}$ is the coherent RP/RN event-probability ratio; $\Delta^{\mathrm{phase}}$ is Eq.~\eqref{eq:phase-interaction}; $\Delta_{\log\Gamma}$ is its mean-$\log\GammaPeak$ counterpart; $R_{\mathrm{alloc}}$ compares top-score with uniform allocation at equal continuation-evaluation counts. Brackets give $95\%$ intervals. Ranges in the text span setting-specific point estimates, not pooled effects.}
\centering
\small
\setlength{\tabcolsep}{4pt}
\begin{ruledtabular}
\begin{tabular}{cccccc}
$n$ & $\delta$ & $R_{\mathrm{RP/RN}}$ & $10^4\Delta^{\mathrm{phase}}$ & $\Delta_{\log\Gamma}$ & $R_{\mathrm{alloc}}$\\
\hline
8 & 1 & $4.82$ [$4.35$, $5.36$] & $35.43$ [$32.77$, $38.17$] & $0.05780$ [$0.05462$, $0.06092$] & $2.09$ [$1.98$, $2.20$]\\
8 & 2 & $4.47$ [$4.11$, $4.88$] & $126.41$ [$118.67$, $134.48$] & $0.07266$ [$0.06843$, $0.07695$] & $2.23$ [$2.15$, $2.31$]\\
8 & 3 & $7.15$ [$6.47$, $7.93$] & $228.70$ [$216.50$, $241.20$] & $0.10229$ [$0.09731$, $0.10726$] & $2.32$ [$2.22$, $2.41$]\\
\noalign{\vskip 3pt}
10 & 1 & $2.58$ [$2.38$, $2.79$] & $54.20$ [$49.15$, $59.32$] & $0.04401$ [$0.04085$, $0.04721$] & $1.90$ [$1.83$, $1.97$]\\
10 & 2 & $3.06$ [$2.82$, $3.32$] & $69.37$ [$63.67$, $75.16$] & $0.05306$ [$0.04974$, $0.05631$] & $1.96$ [$1.88$, $2.03$]\\
10 & 3 & $3.09$ [$2.88$, $3.31$] & $245.24$ [$228.13$, $262.45$] & $0.06679$ [$0.06228$, $0.07110$] & $2.00$ [$1.94$, $2.06$]\\
\noalign{\vskip 3pt}
12 & 1 & $2.64$ [$2.39$, $2.92$] & $16.09$ [$14.31$, $17.93$] & $0.03547$ [$0.03302$, $0.03800$] & $1.89$ [$1.79$, $1.99$]\\
12 & 2 & $2.37$ [$2.19$, $2.57$] & $55.07$ [$49.40$, $60.70$] & $0.04393$ [$0.04063$, $0.04724$] & $1.95$ [$1.88$, $2.02$]\\
12 & 3 & $2.89$ [$2.67$, $3.13$] & $71.97$ [$65.78$, $78.27$] & $0.05259$ [$0.04922$, $0.05599$] & $1.99$ [$1.92$, $2.06$]\\
\noalign{\vskip 3pt}
14 & 1 & $2.01$ [$1.87$, $2.16$] & $32.17$ [$28.63$, $35.71$] & $0.03200$ [$0.02953$, $0.03441$] & $1.79$ [$1.73$, $1.86$]\\
14 & 2 & $2.12$ [$1.97$, $2.27$] & $34.09$ [$30.50$, $37.59$] & $0.03350$ [$0.03110$, $0.03579$] & $1.82$ [$1.75$, $1.89$]\\
14 & 3 & $2.38$ [$2.23$, $2.55$] & $148.02$ [$135.33$, $161.02$] & $0.04936$ [$0.04595$, $0.05272$] & $1.88$ [$1.83$, $1.94$]\\

\end{tabular}
\end{ruledtabular}
\end{table}

\clearpage

\clearpage
\twocolumngrid
\bibliography{apssamp}

\end{document}